\documentclass[12pt,letterpaper]{article}

\usepackage{graphicx,array}
\usepackage{url}
\usepackage{color}
\usepackage{latexsym}
\usepackage{amsthm}
\usepackage{amsmath}
\usepackage{amssymb}
\usepackage{amsfonts}
\usepackage[numbers,sort&compress]{natbib}
\usepackage{bm}
\usepackage{slashed}
\usepackage{mathrsfs}
\usepackage{enumerate}
\usepackage{tikz}
\usepackage{feynmf}
\usepackage{siunitx}
\usepackage{mdframed}
\usepackage{setspace}  
\usepackage{esvect}
\usepackage{esint}
\usepackage{subcaption}
\usepackage[normalem]{ulem}
\usepackage{setspace}
\usepackage{multirow}
\usepackage{diagbox}
\usepackage{soul}
\usepackage{multirow}
\graphicspath{{Charts/}}

\usepackage[version=4]{mhchem}

\usepackage[utf8x]{inputenc}%
\usepackage{tcolorbox}%
\usepackage[normalem]{ulem} 

\usepackage{hyperref} 
\hypersetup{
    colorlinks=true,       
    linkcolor=red,          
    citecolor=blue,        
    filecolor=magenta,      
    urlcolor=blue           
}
\usepackage[all]{hypcap} 

\usepackage{natbib}
\newcommand{\nc}{\newcommand}

\nc{\beq}{\begin{equation}}  
\nc{\eeq}{\end{equation}}  
\nc{\beqa}{\begin{eqnarray}}  
\nc{\eeqa}{\end{eqnarray}}  
\nc{\bit}{\begin{itemize}}  
\nc{\eit}{\end{itemize}}
\DeclareMathOperator\arctanh{arctanh}

\newcommand{\eg}{{\it e.g.}}
\newcommand{\ie}{{\it i.e.}}

\usepackage{floatrow}
\newfloatcommand{capbtabbox}{table}[][\FBwidth]

\usepackage{blindtext}

\title{
{\bf Variant Nelson-Barr Mechanism with \\ Minimal Flavor Violation}
	\author{\large Yang Bai and George N. Wojcik}
	\date{\small \it 
		Department of Physics, University of Wisconsin-Madison, Madison, WI 53706, USA\\
	    }
}

\begin{document}

\maketitle

\setlength{\parskip}{0.2ex}

\begin{abstract}	
Within the general framework of using spontaneous CP violation to solve the strong CP problem, we construct a variant Nelson-Barr model in which the Standard Model (SM) quark contribution to the strong CP phase is cancelled by new heavy QCD-charged fermions. This cancellation is ensured by choosing conjugate representations for the new colored states under the same global flavor symmetry of SM quarks. Choosing the global flavor symmetry to be that of minimal flavor violation, we suppress higher-order corrections to the strong CP phase to well below current experimental constraints. More than two dozen massless Goldstone bosons emerge from spontaneous flavor symmetry breaking, which yield strong astrophysical constraints on the symmetry breaking scale. In the early universe, the Goldstone bosons can be thermally produced from their interactions with the heavy colored fermions and contribute to $\Delta N_{\rm eff}$ at a measurable level. As a function of reheating temperature, the predicted $\Delta N_{\rm eff}$ shows an interesting plateau behavior we dub the ``flavor stairway", which encodes information about the SM quark flavor structure. 
\end{abstract}

\thispagestyle{empty}  
\newpage  
  
\setcounter{page}{1}  

\begingroup
\hypersetup{linkcolor=black,linktocpage}
\tableofcontents
\endgroup
\newpage

\section{Introduction}\label{sec:Introduction}
The strong CP problem \cite{tHooft:1976rip} represents a major ongoing puzzle in physics: harsh experimental limits on the neutron electric dipole moment \cite{Baker:2006ts,Pendlebury:2015lrz,PhysRevLett.116.161601} lead to draconian constraints on the strong CP phase
of $\overline{\theta} \lesssim O(10^{-10})$. The most well-studied solution to the strong CP problem is to introduce a global $U(1)$ symmetry, the Peccei-Quinn (PQ) symmetry~\cite{Peccei:1977hh,Peccei:1977ur}, with its corresponding Goldstone boson, the axion~\cite{Weinberg:1977ma,Wilczek:1977pj}, to dynamically relax $\overline{\theta}$ to zero. 
On the other hand, models without a PQ symmetry are also viable: a promising variety of solutions have been proposed in frameworks using spontaneous CP breaking, where the CP is a good symmetry in the UV theory and the weak CP-violating angle is obtained after spontaneous CP breaking~\cite{Nelson:1983zb,Nelson:1984hg,Barr:1984qx,Hiller:2001qg,Harnik:2004su,Vecchi:2014hpa,Dine:2015jga}. These classes of solutions share a common conceptual approach: for a theory with sets of fermions $\psi_r$ in representation $r$ under the QCD $SU(3)_c$ group [note that the Standard Model (SM) quarks must be contained in $\psi_{\mathbf{3}}$], the general strong CP phase $\overline{\theta}$ is given by
%
$\overline{\theta} = \theta + 2\sum_r C(r)\, \mbox{arg}[\mbox{det}(\mathcal{M}(\psi_r))]$, 
%
where $\theta/(32 \pi^2)$ is the coefficient of the term $G \widetilde{G}$ in the QCD action (where $G$ is the gluon field strength tensor and $\widetilde{G}$ is its dual), while $C(r)$ is the Dynkin index of the representation $r$ (normalized so that $C(\mathbf{3}) = 1/2$) and $\mathcal{M}(\psi_r)$ is the mass matrix for the collection of fermions $\psi_r$. If CP is a good symmetry of the UV theory, then $\theta=0$, and a model resolves the strong CP problem if the QCD-charged fermion mass matrices satisfy
%
\beqa
\label{eq:general-cond}
\sum_r C(r)\, \mbox{arg}\Bigl[\mbox{det}\bigl(\mathcal{M}(\psi_r)\bigr)\Bigr] = 0 ~,
\eeqa
or at least that said quantity is small enough to satisfy current experimental constraints on $\overline{\theta}$. Because the SM contains an $O(1)$ CP-violating phase in the weak CKM matrix, including the SM fermion content in the sum of Eq.~\eqref{eq:general-cond} while still satisfying the condition for a vanishing $\overline{\theta}$ is somewhat non-trivial. Nelson-Barr models \cite{Nelson:1983zb,Nelson:1984hg,Barr:1984qx} are a well-known and (at least in its minimal realization) simple construction which satisfies the condition in \eqref{eq:general-cond}. These models realize \eqref{eq:general-cond} by imposing CP symmetry and introducing heavy vector-like quarks. CP is spontaneously broken via the vacuum expectation value (VEV) of a complex scalar which couples the SM and vector-like quarks, introducing an $O(1)$ weak CP phase but preserving the strong CP phase at tree level. The minimal Nelson-Barr model generally suffers from significant challenges, notably that the tree-level condition for a vanishing strong CP phase is generally unstable against corrections from higher-dimensional operators and radiative corrections \cite{Dine:2015jga,Perez:2020dbw}. There have been a variety of approaches which extend the minimal Nelson-Barr framework in order to address these difficulties, for example by supersymmetry \cite{Fujikura:2022sot,Evans:2020vil,Hiller:2001qg}, extra dimensions \cite{Cheung:2007bu,Girmohanta:2022giy}, or compositeness \cite{Perez:2020dbw,Valenti:2021xjp}.

A common feature of the Nelson-Barr-like constructions we have listed above, as well as many other spontaneous CP-violating solutions to the strong CP problem \cite{Vecchi:2014hpa,Diaz-Cruz:2016pmm}, are constructed so that the SM quark mass matrices alone possess no overall complex phase at leading order, that is, the strong CP problem can be said to be solved from the ``bottom up'': After integrating out whatever heavier-scale physics might exist, the contribution of the SM quarks to Eq.~(\ref{eq:general-cond}) is small enough that its contribution to the strong CP phase is small enough to not run afoul of experimental constraints on its own. While this is a compelling condition, the SM quark mass matrices can generally enjoy a large complex phase in their determinants while Eq.~(\ref{eq:general-cond}) is still satisfied: the strong CP phase is sensitive to physics at arbitrary scales, so a complex SM quark mass matrix determinant is allowed as long as heavy color-charged fermions cancel that complex phase in the UV-complete theory. This is generally the mechanism behind solutions to the strong CP problem involving parity symmetry~\cite{Mohapatra:1978fy,Beg:1978mt,Babu:1989rb,Barr:1991qx,Dunsky:2019api}, in which heavy mirror fermions cancel the strong CP phase contribution of the SM quarks (but see Ref.~\cite{deVries:2021pzl} for a detailed investigation of the potentially sizable two-loop contribution).

In this work, we present a spontaneous CP-violation solution to the strong CP problem which takes a similar approach. In our construction, a flavor symmetry enforces a relationship between the SM quark mass matrices and those of much heavier color-charged fermions, so that the overall strong CP phase is zero. We explore a comparatively straightforward realization of this paradigm, in which the global flavor symmetry is that of minimal flavor violation (MFV) in the quark sector~\cite{Chivukula:1987py,DAmbrosio:2002vsn}, and we assume that the flavor and CP symmetries are spontaneously broken at a high scale by two scalars which impart mass matrices to the SM quarks consistent with the observed pattern of quark masses and mixings. Heavy vector-like quarks then acquire masses through the same scalars as the SM quarks, but have conjugate representations under the flavor group, ensuring that a complex phase of the SM quark mass matrix determinants is cancelled by an opposite contribution from the heavy vector-like quarks. Our use of the MFV paradigm, leading to the weak CP phase being the only physical phase in the quark mass matrices, insulates $\overline{\theta}$ from large corrections from higher-dimensional operators, and we find estimated leading quantum corrections to the strong CP phase of $\lesssim O[10^{-(13-24)}]$. Since this estimate does not include further suppression from, \eg, loop factors, we therefore find that this construction offers significantly greater suppression of the strong CP phase than other models which employ a flavor symmetry to enforce the condition of Eq.~(\ref{eq:general-cond}) \emph{without} phase cancellation from vector-like quarks, \eg ~\cite{Vecchi:2014hpa}. We further find that, in contrast to most parity- and CP-based models to address the strong CP problem, imposing a global flavor symmetry provides intriguing low-energy phenomenology signatures, akin to those stemming from flavorful axion models~\cite{Arias-Aragon:2017eww,Green:2021hjh,Baumann:2016wac,delaVega:2021ugs,Bauer:2021mvw}.

Our paper is laid out as follows. In Section \ref{sec:basic-model}, we outline the model and demonstrate that the model resolves the strong CP problem at tree-level, and that quantum corrections to the strong CP phase are within current experimental limits. In Section \ref{sec:GB}, we discuss the most distinctive phenomenological feature of the model, the large number of flavorful Goldstone bosons (GB's) emerging from spontaneous breaking of the global MFV symmetry. In Section \ref{sec:pheno}, we discuss in some detail the phenomenological signatures of these GB's in the simplest case, namely that MFV is treated as a true global symmetry and all Goldstone bosons remain massless. Finally, in Section \ref{sec:conclusion}, we discuss our results and directions for future work on the general class of constructions in which a flavor symmetry enforces the cancellation between complex phases in the determinant of the SM quark mass matrices and those of some heavier QCD-charged fermions. In Appendix \ref{sec:more-operators}, we present a detailed proof that the solution to the strong CP problem is protected from high-order corrections. We provide a detailed calculation for the GB thermal history in Appendix \ref{sec:thermal-GB} and a renormalizable model with an analogous setup in Appendix \ref{sec:ren-model}.

\section{Basic Model}\label{sec:basic-model}
As a brief review and to present a point of comparison to our construction, we remind the reader of the basic construction of Nelson-Barr models. In these constructions, the fermion mass matrix (including both SM and exotic quarks) is taken to be $\mathcal{M} = \mbox{diag}\{\mathcal{M}^u, \mathcal{M}^d\}$, where $\mathcal{M}^{u, d}$ are the up-like and down-like quark mass matrices, in a block-diagonal form
\beqa
\mathcal{M}^{u, d} \propto
\begin{pmatrix}
A & B \\
0 & C 
\end{pmatrix} ~, 
\eeqa
with $A^\dagger = A$ and $C^\dagger = C$ such that $\mbox{arg}[\mbox{det}(\mathcal{M}^{u, d})] = 0$. The ``$0$" matrix is enforced by imposing some discrete or continuous symmetries. 

As a variant of the Nelson-Barr Mechanism, we explore another class of models with the general structure of fermion matrices as
\beqa
\label{eq:general-matrix}
\mathcal{M} \propto
\begin{pmatrix}
r_d\, A_d & 0  & 0 & 0 \\
0 & r_u\, A_u  & 0 & 0 \\
0 & 0 &  A_d^* & 0  \\ 
0 & 0 & 0 &  A_u^* 
\end{pmatrix} ~, 
\eeqa
with $r_{d, u}$ as real numbers and the product $r_u r_d >0$, and $A_{d, u}$ as complex $3\times 3$ matrices. This mass matrix structure also satisfies the general condition in \eqref{eq:general-cond}. The ordinary quark mass matrix is determined by the upper two blocks, $r A$, while new vector-like fermions have mass matrices in the lower two blocks, $A^*$. To impose this non-trivial relation between the upper and lower block matrices, we impose the global flavor symmetry of the SM quarks $U(3)_{q_L} \times U(3)_{d_R} \times U(3)_{u_R}$ (as in the MFV) in the limit of vanishing Yukawa couplings. By assigning conjugate representations under flavor symmetry for the new vector-like fermions, the structure with $A$ and $A^*$ in \eqref{eq:general-matrix} can be enforced.~\footnote{Another possibility, with parity-reversed representations for new fermions and $A^*$ replaced by $A^\dagger$, will achieve the same goal with similar phenomenology. Constructions also exist with fewer heavy vector-like fermions. For example, a model with only one up-like and one down-like vector-like quark will satisfy Eq.~(\ref{eq:general-cond}) if their mass terms are proportional to $\det A^*_{u}$ and $\det A^*_{d}$, which can be enforced by properly assigning $U(1)$ charges.} In contrast to MFV, where spurion fields are introduced to track the flavor symmetry breaking effects, we promote the spurion fields to dynamical fields and assume that some potential of the fields spontaneously breaks the flavor and CP symmetries. For the three $U(1)$'s in the global flavor symmetry and leaving aside the baryon number symmetry, we will keep the other two as a good symmetry in the model and denote them as $U(1)_d$ and $U(1)_u$, respectively. The model particle content and representations under both SM gauge symmetries and the flavor symmetries are listed in Table~\ref{tab:model}.

\begin{table}[hb!]
	\centering
	\renewcommand{\arraystretch}{1.4}
		\begin{tabular}{ c | c | c | c | c |c |c }
			\hline \hline
  &   $[SU(3)_c \times SU(2)_W \times U(1)_Y] $ & $SU(3)_{q_L}$ & $SU(3)_{d_R}$ & $SU(3)_{u_R}$ & $U(1)_u$ & $U(1)_d$ \\ \hline
  $q_L$ & $(3, 2)_{1/6}$ & 3 & 1 & 1 & 0 &  0 \\ \hline 
  $d_R$ & $(3, 1)_{-1/3}$ & 1 & 3 & 1 & 0 & +1 \\ \hline 
  $u_R$ & $(3, 1)_{2/3}$ & 1 & 1 & 3 & +1 & 0 \\ \hline \hline
  $B_L$ & $(3, 1)_{Q_B}$ & $\overline{3}$ & 1 & 1 & 0 & 0 \\ \hline
  $B_R$ & $(3, 1)_{Q_B}$ & 1 & $\overline{3}$ & 1 & 0 & -1 \\ \hline  
  $T_L$ & $(3, 1)_{Q_T}$ & $\overline{3}$ & 1 & 1 & 0 & 0 \\ \hline
  $T_R$ & $(3, 1)_{Q_T}$ & 1 & 1 & $\overline{3}$ & -1 & 0 \\ \hline  \hline
   $H$ & $(1, 2)_{1/2}$ & 1 & 1 & 1 & 0 &  0 \\ \hline 
  $\Sigma_d$ & $(1, 1)_0$ & 3 & $\overline{3}$ & 1  & 0 & -1 \\ \hline  
  $\Sigma_u$ & $(1, 1)_0$ & 3 & 1 & $\overline{3}$ & -1 & 0 \\ \hline  
		 \hline 
		\end{tabular}
	\caption{
	Matter content of the model. The SM quarks are denoted by $q_L$, $d_R$, and $u_R$ and arranged in triplets of the MFV non-Abelian flavor group. $B_{L, R}$ and $T_{L, R}$ are vector-like down-like and up-like quarks arranged in MFV anti triplets. The hypercharges $Q_B$ and $Q_T$ with $Q_B \neq Q_T$ are kept as general parameters with $Q_B = -1/3$ and $Q_T = +2/3$ to have new fermions match the electric charges of down-type and up-type quarks. The last three rows contain the scalar particles with $\Sigma_u$ and $\Sigma_d$ as SM gauge singlets that spontaneously break the MFV group and CP. All global flavor symmetries do not have a mixed anomaly with $SU(3)_c$.
		} \label{tab:model}
\end{table}

With the symmetry and matter content, the leading operators containing fermion fields up to the dimension-five level are 
\beqa \label{eq:fermion-mass-terms}
\mathcal{L}_F \supset - y_d \,H\, \frac{\overline{q}_L\,\Sigma_d\,d_R}{\Lambda} - \eta_d \, \overline{B}_L\,\Sigma_d^*\, B_R  - y_u\, \widetilde{H}\, \frac{\overline{q}_L\,\Sigma_u\,u_R}{\Lambda} - \eta_u \,  \overline{T}_L\,\Sigma_u^* \, T_R \,+\, h.c.
\eeqa
Here, $\widetilde{H} \equiv i\,\sigma_2 H^*$. Under P transformation, one has $\overline{\psi}_{L, i} \chi_{R, j}  \leftrightarrow \overline{\psi}_{R, i} \chi_{L, j}$. Under C transformation, one has $H \leftrightarrow H^*$, $\Sigma_{d, u} \leftrightarrow \Sigma_{d, u}^*$ and $\overline{\psi}_{L,i} \psi_{R,j} \leftrightarrow \overline{\psi}_{L,j} \psi_{R,i}$. To conserve CP symmetry, all Yukawa couplings $y_{d, u}$ and $\eta_{d, u}$ are required to be real. Note that because chiral representations for SM quarks under electroweak gauge symmetry, the Lagrangian does not conserve either P or C, but conserves CP. 

Other than the interactions including fermions, we also have a potential for the scalar fields to spontaneously break the global flavor symmetry
\beqa
\label{eq:scalar-potential}
V(\Sigma_d, \Sigma_u) = V_{\rm symm.}(\Sigma_d, \Sigma_u) + V_{\rm soft}(\Sigma_d, \Sigma_u) ~,
\eeqa
which contains both symmetric and soft-symmetry-breaking potential terms as well as other scalar fields to achieve the particular spontaneous symmetry-breaking pattern. For some detailed studies to minimize the potential and identify the vacua, see for example Refs.~\cite{Alonso:2011yg,Espinosa:2012uu,Fong:2013dnk}. After symmetry breaking, the VEV's of $\Sigma_d$ and $\Sigma_u$ are proportional to down-type and up-type quark mass matrices or $\langle\Sigma_d\rangle \propto M_d$ and $\langle\Sigma_u\rangle \propto M_u$. 

Separating fermions into two groups, $(d, B)$ and $(u, T)$, and after electroweak symmetry breaking with $\langle H \rangle^T = (0, v/\sqrt{2})$ with $v=246$\,GeV, we have their mass matrices as 
\beqa
\renewcommand{\arraystretch}{1.3}
\label{eq:mass-matrix}
\mathcal{M}^d = 
\begin{pmatrix}
\dfrac{y_d\,v}{\sqrt{2}}\, \dfrac{\langle \Sigma_d \rangle}{\Lambda} & 0 \\
0 &  \eta_d \, \langle \Sigma_d \rangle^* 
\end{pmatrix} ~, \qquad \qquad
\mathcal{M}^u = 
\begin{pmatrix}
\dfrac{y_u\,v}{\sqrt{2}}\, \dfrac{\langle \Sigma_u \rangle}{\Lambda} & 0 \\
0 &  \eta_u \, \langle \Sigma_u \rangle^* 
\end{pmatrix} ~.
\eeqa
The mass matrices satisfy the condition of $\mbox{arg}[\mbox{det}(\mathcal{M}^{d} \mathcal{M}^{u})] = 0$ (as long as the product $y_u y_d \eta_u \eta_d > 0$), so the strong CP problem is solved, at least based on the leading-order operators in \eqref{eq:fermion-mass-terms}.

\subsection{Corrections to the Mass Matrix Structure}
Having established the leading-order validity of our model, we can address its stability under higher-order corrections. Notably, the MFV paradigm (specifically that all flavor violating effects in the model come from the CKM matrix and quark mass ratios) greatly suppresses the quantum corrections to the strong CP phase---since the only flavor violation occurs in the scalars $\Sigma_{u,d}$ and is governed by the SM Yukawa structure, quantum corrections are harshly suppressed by CKM factors.~\footnote{Since we are working with a continuous global symmetry, a reader may be concerned that the strong CP phase will be subject to large corrections due to Planck-scale explicit symmetry breaking terms, as occurs in axion models \cite{Barr:1992qq}. In contrast to the case of the axion and the Peccei-Quinn symmetry, however, we note that the global flavor group we are considering is non-anomalous (up to mixed anomalies with the SM electroweak symmetry, which can be easily resolved by introducing QCD-singlet fermions with nontrivial representations under the flavor group). Hence, we can avoid these draconian corrections from Planck-scale operators by, for example, assuming that our global symmetry is actually a very feebly-coupled local symmetry.} We can estimate this by considering higher-dimensional operators that might emerge, for example, from radiative corrections. Inspecting \eqref{eq:mass-matrix}, we can see that there are two classes of possibly problematic contributions: non-trivial contributions to the off-diagonal blocks in \eqref{eq:mass-matrix}, and contributions to the diagonal blocks which spoil the phase relationship between the SM and vector-like quark mass matrices. We discuss both possible corrections in turn. 

For the off-diagonal entries, one can show that there is not any correction---in other words, its zeros are protected by the symmetry of the model. For some hypercharge choices of $(Q_B, Q_T)$, the SM electroweak gauge symmetry can forbid some operators. Even independent of the values of  $(Q_B, Q_T)$, one can show that the global non-Abelian flavor symmetry $SU(3)_{q_L} \times SU(3)_{d_R} \times SU(3)_{u_R}$ can protect any corrections to the off-diagonal entries. Using the upper off-diagonal entry of $\mathcal{M}^d$ as an example, its representation under the non-Abelian flavor symmetry is $(3, 3, 1)$. If there is one operator containing $x$ $\Sigma_d$, $y$ $\Sigma_d^*$, $z$ $\Sigma_u$ and $\omega$ $\Sigma_u^*$ with $x, y, z, \omega \in \mathcal{Z}$, one has the total numbers of Young tableaux boxes as $x + 2 y + z + 2 \omega$ for $SU(3)_{q_L}$, $2 x + y$ for $SU(3)_{d_R}$ and $2 z + \omega$ for $SU(3)_{u_R}$, respectively. To match the representation of $(3, 3, 1)$, one need to have $x + 2 y + z + 2 \omega = 1 \bmod 3$, $2 x + y = 1 \bmod 3$ and $2 z + \omega = 0 \bmod 3$. Twice of the last two relations minus the first relation provides $3 x + 3 z  = 1 \bmod 3$, which is false and completes the proof. For other off-diagonal entries in $\mathcal{M}^d$ and $\mathcal{M}^u$, one can have a similar proof for their zero values. 

For the diagonal entries, additional higher-dimensional operators could break the relation between light quark and heavy quark mass matrices. Preserving the global symmetry of the model, one leading operator contributing to the strong CP phase happens at the dimension-13 level (assuming our construction is valid up to a UV cutoff of $\sim \Lambda$, the same scale that appears in the dimension-5 operator imparting the SM quark masses; see Appendix~\ref{sec:more-operators} for details) and is 
\beqa\label{eq:leading-CP-op}
\zeta_u\,\widetilde{H}\, \frac{\overline{q}_L\, (\Sigma_d \Sigma_d^\dagger)^2 (\Sigma_u \Sigma_u^\dagger)^2 (\Sigma_d \Sigma_d^\dagger)(\Sigma_u \Sigma_u^\dagger)\, \Sigma_u\,u_R}{\Lambda^{13}} ~,
\eeqa
which serves as a correction to the SM quark mass matrix. Defining the correction matrix $\mathcal{C} \equiv y_u^{-1} \zeta_u (\Sigma_d \Sigma_d^\dagger)^2 (\Sigma_u \Sigma_u^\dagger)^2 (\Sigma_d \Sigma_d^\dagger)(\Sigma_u \Sigma_u^\dagger)/\Lambda^{12}$, we can then express our results in terms of CKM matrix parameters and the quark masses. To extract numerical results, we specify our quark masses by setting  $m_t$, the top quark mass, at its $\overline{\mbox{MS}}$ value, and evaluate all other quark masses at the scale $m_t$ in the $\overline{\mbox{MS}}$ scheme, leading to
\beqa \label{eq:quark-masses}
    \begin{matrix}
        m_u = 1.21 \; \textrm{MeV}\,, & m_c = 604 \; \textrm{MeV}\,, & m_t = 163 \; \textrm{GeV}\,, \\
        m_d = 2.58 \; \textrm{MeV}\,, & m_s = 52.7 \; \textrm{MeV}\,, & m_b = 2.73 \; \textrm{GeV} \,.
    \end{matrix}
\eeqa
Since quark mass ratios are insensitive to QCD running, the specific scale at which we evaluate our masses has a negligible effect on most of our results.

Equipped with definite values for the quark masses, we find that the nonzero contribution to the argument of the determinant from our operator in Eq.~(\ref{eq:leading-CP-op}) is
\beqa\label{eq:leading-CP-op-numerics}
\mbox{arg}[\mbox{det}(\mathcal{M}^u )] &=& \mbox{arg}[ \mbox{det}(\mathbb{I} + \mathcal{C} ) ] \approx \mbox{Im}[\mbox{Tr}\,\mathcal{C}]  \nonumber \\
&\approx& 2 \,|J|\, y_u^{-1} \zeta_u\,\left(\frac{\sqrt{2}\,m_t}{y_u\,v}\right)^6  \left(\frac{\sqrt{2}\,m_t}{y_d\,v}\right)^6 \,  \left(\frac{m_b}{m_t}\right)^4\,\left(\frac{m_c}{m_t}\right)^2 \left(\frac{m_s}{m_t}\right)^2 ~,
\eeqa
where the Jarlskog invariant $J = \cos{\theta_{12}} \cos^2{\theta_{13}} \cos{\theta_{23}} \sin{\theta_{12}} \sin{\theta_{13}} \sin{\theta_{23}} \sin{\delta}$ in terms of CKM mixing angles and Dirac CP phase~\cite{Jarlskog:1985ht}. Here, we have kept the leading term in quark mass ratios, which agrees with the mass ratio dependence just based on SM interactions, in Ref.~\cite{Ellis:1978hq}. Numerically, we have the result $\sim 3.1\times 10^{-13}$ for $y_u \sim 1$, $y_d \sim m_b/m_t$ and $\zeta_u \sim 1$, which is below the current experimental bound. Given that the coupling $\zeta_u$ could contain many powers of a loop factor $1/(16 \pi^2)$, the additional corrections for the fermion mass matrices are negligible and the solution to the strong CP problem in this model stays valid. For a more thorough discussion of other higher-dimensional operator contributions to the strong CP phase, see Appendix~\ref{sec:more-operators} for detail. 

\section{MFV Goldstone Bosons}
\label{sec:GB}

Because of the spontaneous breaking of the global flavor symmetry, massless (or approximately massless if there are additional explicit global symmetry breaking terms), Goldstone bosons are the leading low energy prediction for the model here. For non-Abelian global symmetries, there are $3\times 8 = 24$ GB's. For the Abelian ones, both $U(1)_u$ and $U(1)_d$ are orthogonal to $U(1)_Y$ and the baryon-number $U(1)_B$ symmetry and are also spontaneously broken, leading to 2 more GB's. 
Altogether, there are 26 GB's in the low-energy effective theory. Note that the 24 non-Abelian GB's are intrinsic to the quark sector in the MFV framework. The 2 Abelian GB's are model dependent and required for solving the strong CP problem in our model. 

Using the non-linear parametrization, we have the GB degrees of freedom as
\beqa \label{eq:sigma-defs}
\Sigma_d  
&\equiv& f_d \,e^{i\,t^a\,\theta_q^a/f_u}\, R_d\, e^{- i \,t^{\tilde{a}}\,\theta_d^{\tilde{a}}/f_d}
~, \\
 \Sigma_u  
&\equiv& f_u\, e^{i\,t^a\,\theta_q^a/f_u}\,V^\dagger \,R_u\, e^{- i \,t^{\tilde{a}}\,\theta_u^{\tilde{a}}/f_u}
~, \nonumber
\eeqa
where $R_d = \mbox{diag}(m_d/m_b, m_s/m_b, 1)$ and $R_u = \mbox{diag}(m_u/m_t, m_c/m_t, 1)$ and $V = V_{\rm CKM}$ is the CKM unitary matrix. Here, $a = 1, \cdots, 8$. ${\tilde{a}} = 0,1,\cdots, 8$ with $t^0 = \mathbb{I}_3/\sqrt{6}$. The normalization conventions for all generators are $\mbox{tr}[ t^a t^b] = \frac{1}{2}\delta^{ab}$ and $\mbox{tr}[ t^{\tilde{a}} t^{\tilde{b}}] = \frac{1}{2}\delta^{{\tilde{a}}{\tilde{b}}}$. The relationship between the scales $f_u$ and $f_d$ and the scale $\Lambda$ in, \eg,   Eq.~(\ref{eq:fermion-mass-terms}), is also apparent in Eq.~(\ref{eq:sigma-defs})---in order to maintain a top quark Yukawa coupling of $O(1)$, we shall need $\Lambda \sim f_u$ (as we shall see in our discussion of a renormalizable version of this model in Appendix \ref{sec:ren-model}). The specific value of $\Lambda$ has no relevance to our discussion, since it can be absorbed into a redefinition of the SM up-like Yukawa coupling $y_u$, and so we set $\Lambda = f_u$ in our later discussion. 

As it will be significant for our phenomenology later, it is useful to remark briefly on the relative magnitudes of the two dimensionful scales $f_d$ and $f_u$ in Eq.~\eqref{eq:sigma-defs}. Referring to Eq.~\eqref{eq:fermion-mass-terms}, it is straightforward to derive that $f_d/f_u = y_u\,m_b/(y_d\,m_t)$ where $y_u$ and $y_d$ are the Yukawa coupling terms appearing in Eq.~(\ref{eq:fermion-mass-terms}). There are two possible interesting regimes of values in which both $y_u$ and $y_d$ remain perturbative: either $f_d \sim f_u$ or $f_d \sim m_b f_u/m_t \ll f_u$. If $f_d \sim f_u$, then the large mass hierarchy between $t$ and $b$ quarks is achieved by having $y_d \ll y_u$, while if $f_d \sim m_b f_u/m_t$, then the Yukawa couplings $y_u$ and $y_d$ are approximately equal.

The kinetic terms of GBs from $\Sigma_d$ and $\Sigma_u$ are
\beqa \label{eq:goldstone-kinetic-terms}
\mathcal{L} 
&\supset& H_1^{ab}\, \partial_\mu \theta^a_q \, \partial^\mu \theta^b_q  + H_2^{\tilde{a}\tilde{b}}\, \partial_\mu \theta^{\tilde{a}}_d \, \partial^\mu \theta^{\tilde{b}}_d
- H_3^{a\tilde{b}}\, \partial_\mu \theta^a_q \, \partial^\mu \theta^{\tilde{b}}_d  - H_3^{T\,\tilde{a}b}\, \partial_\mu \theta^{\tilde{a}}_d \, \partial^\mu \theta^b_q
\nonumber \\
&&+\,  K_1^{ab}\, \partial_\mu \theta^a_q \, \partial^\mu \theta^b_q  + K_2^{\tilde{a}\tilde{b}}\, \partial_\mu \theta^{\tilde{a}}_u \, \partial^\mu \theta^{\tilde{b}}_u
- K_3^{a\tilde{b}}\, \partial_\mu \theta^a_q \, \partial^\mu \theta^{\tilde{b}}_u  - K_3^{T\,\tilde{a}b}\, \partial_\mu \theta^{\tilde{a}}_u \, \partial^\mu \theta^b_q  ~.
\eeqa
Here, 
\beqa
&&H_1^{ab} \equiv \frac{f_d^2}{f_u^2}\, \mbox{Tr} [ t^a R_d R_d t^b ] \,, \quad H_2^{\tilde{a}\tilde{b}} \equiv \mbox{Tr} [ t^{\tilde{a}} R_d R_d t^{\tilde{b}} ] \,, \quad H_3^{a\tilde{b}} \equiv \frac{f_d}{f_u}\,\mbox{Tr} [ t^a R_d t^{\tilde{b}} R_d  ]  \,,\nonumber \\
&&
K_1^{ab} \equiv \mbox{Tr} [ t^a V^\dagger R_u R_u V t^b ] \,, \quad K_2^{\tilde{a}\tilde{b}} \equiv \mbox{Tr} [ t^{\tilde{a}} R_u R_u t^{\tilde{b}} ] \,, \quad K_3^{a\tilde{b}} \equiv \mbox{Tr} [ t^a V^\dagger R_u t^{\tilde{b}} R_u V] ~.
\eeqa
In general, one need to diagonalize the $26\times 26$ kinetic mixing matrix to obtain orthonormal states. Rotating the fermion fields to the mass eigenstates, one can derive the single GB's couplings to fermions (ignoring the explicitly symmetry-breaking parameters) as
\beqa
\label{eq:GB-couplings}
\mathcal{L} 
&\supset& - \frac{1}{f_u}\, \overline{d}_L \gamma^\mu t^a\, (\partial_\mu \theta_q^a)\,d_L + \frac{1}{f_u}\,\overline{B}_L \gamma^\mu\,t^{a *} \, (\partial_\mu \theta_q^a)\,B_L  \nonumber \\
&&- \frac{1}{f_d}\, \overline{d}_R \gamma^\mu t^{\tilde{a}}\,(\partial_\mu \theta_d^{\tilde{a}}) \, d_R + \frac{1}{f_d}\, \overline{B}_R \gamma^\mu t^{\tilde{a}*}\,(\partial_\mu \theta_d^{\tilde{a}}) \, B_R  \nonumber \\
&& - \frac{1}{f_u}\,\overline{u}_L \gamma^\mu \,V\, t^a\, V^\dagger\, (\partial_\mu \theta_q^a) \, u_L + \frac{1}{f_u}\,\overline{T}_L \gamma^\mu\,V^*\, t^{a*}\,V^T\,(\partial_\mu \theta_q^a) \, T_L  \nonumber \\
&& - \frac{1}{f_u}\, \overline{u}_R \gamma^\mu t^{\tilde{a}}\,(\partial_\mu \theta_u^{\tilde{a}}) \, u_R + \frac{1}{f_u}\,\overline{T}_R \gamma^\mu t^{\tilde{a}*}\,(\partial_\mu \theta_u^{\tilde{a}}) \, T_R ~.
\eeqa
Rotating/rescaling the GBs to the canonically-normalized kinetic basis, their couplings to fermions can be easily derived, albeit with complicated expressions.  

Because of the sequential breaking of the global symmetries, the two GB's with smallest decay constants (for $f_u \sim f_d$) can be identified as the last sequential step of $U(3)_R \rightarrow U(2)_R \rightarrow U(1)_R \rightarrow \emptyset$. In terms of previous fields, they are
\beqa
\label{eq:leading-2-GB}
a_d &\approx& \frac{m_d}{m_b}\,\left(\frac{1}{\sqrt{3}} \theta_d^0 + \frac{1}{\sqrt{2}} \theta_d^3 + \frac{1}{\sqrt{6}} \theta_d^8\right) ~, \nonumber \\
a_u &\approx& \frac{m_u}{m_t}\,\left(\frac{1}{\sqrt{3}} \theta_u^0 + \frac{1}{\sqrt{2}} \theta_u^3 + \frac{1}{\sqrt{6}} \theta_u^8\right) ~,
\eeqa
at the leading order in an expansion parameter of $\lambda \approx 0.22$ (we express all quark mass ratios and CKM mixing matrix entries in terms of powers of $\lambda$ and other order-one numbers). Their couplings to fermions are 
\beqa\label{eq:alpha-couplings}
\mathcal{L} &\supset& - \frac{1}{f_{a_d}} \, \overline{d}_R \gamma^\mu\, t^{a_d} \, (\partial_\mu a_d)\,d_R  
+ \frac{1}{f_{a_d}} \, \overline{B}_R \gamma^\mu\, (\partial_\mu a_d)\, t^{a_d} \,B_R \nonumber \\
&& - \frac{1}{f_{a_u}} \, \overline{u}_R \gamma^\mu\, t^{a_u} \, (\partial_\mu a_u)\,u_R
+ \frac{1}{f_{a_u}} \, \overline{T}_R \gamma^\mu\,(\partial_\mu a_u)\, t^{a_u} \, T_R
 ~,
\eeqa
with $f_{a_d} \equiv f_d\times m_d/m_b$ and $f_{a_u} \equiv f_u\times m_u/m_t$. The corresponding generators are $t^{a_d} = t^{a_u} \equiv \mbox{diag}(1, 0, 0)/\sqrt{2}$. Note that their couplings are mostly ``right handed". This is because of the non-zero CKM matrix: even when $m_u = m_d = 0$, the $U(1)$ symmetry related to the first generation of the left-handed transformation is still broken. The couplings for all 26 GB's to fermions, as a function of $f_u$ and $f_d$, can be derived in the similar manner, and we shall keep them in some of our later phenomenological studies.

Apart from the matter couplings, the absence of a mixed QCD anomaly with the flavor group we have proposed immediately precludes the presence of a coupling between the GB's and the gluons, in contrast to the QCD axion. However, depending on the hypercharge assignments of the heavy vector-like quarks, there may be a non-vanishing mixed anomaly between the $U(1)_d$ and $U(1)_u$ groups and SM hypercharge, which will lead to a GB coupling to photons. In terms of the original GB's defined in Eq.~\eqref{eq:sigma-defs}, these photon couplings are 
\beqa \label{eq:theta-photon-coupling}
\mathcal{L} \supset \bigg\{ \frac{\theta^{0}_u}{f_u}\,3 \sqrt{\frac{3}{2}} \bigg[Q_T^2 - \bigg( \frac{2}{3}\bigg)^2 \bigg] + \frac{\theta^{0}_d}{f_d}\,3\sqrt{\frac{3}{2}} \bigg[Q_B^2 - \bigg( \frac{1}{3}\bigg)^2 \bigg] \bigg\} \frac{\alpha_{\rm em}}{4 \pi} F^{\mu \nu} \widetilde{F}_{\mu \nu}~,
\eeqa
where $Q_B$ and $Q_T$ are the SM hypercharge of the heavy vector-like $B$ and $T$ quarks, respectively, while $\alpha_{\rm em}$ is the usual fine structure constant, $F_{\mu \nu}$ is the electromagnetic field strength tensor, and $\widetilde{F}^{\mu \nu} = \epsilon^{\mu \nu \alpha \beta} F_{\alpha \beta}/2$, with $\epsilon^{0123}=+1$. From Eq.~\eqref{eq:theta-photon-coupling}, we can then straightforwardly find the photon couplings for each canonically normalized GB. For $a_u$ and $a_d$, these coupling terms are
\beqa
\mathcal{L} \supset \bigg\{ \frac{a_u}{f_{a_u}} \frac{3}{\sqrt{2}} \bigg[ Q_T^2 - \bigg( \frac{2}{3}\bigg)^2\bigg] + \frac{a_d}{f_{a_d}} \frac{3}{\sqrt{2}} \bigg[ Q_B^2 - \bigg( \frac{1}{3}\bigg)^2\bigg] \bigg\} \frac{\alpha_{\rm em}}{4 \pi} F^{\mu \nu} \widetilde{F}_{\mu \nu}~.
\eeqa

Before moving on to the phenomenology, it is useful to briefly discuss what coupling terms may be phenomenologically significant in our model. In particular, although the derivative couplings in Eqs.~\eqref{eq:GB-couplings} and \eqref{eq:alpha-couplings} contain both vector and axial-vector currents, flavor-diagonal vector GB couplings do not enter our phenomenology. This can be explicitly seen via integrating the action by parts and applying the classical equations of motion, after which flavor-diagonal vector couplings vanish---it also can be expected from vector current conservation~\cite{Baumann:2016wac}.

So, to summarize, our GB's experience three types of couplings to SM fields: flavor-diagonal axial-vector couplings, flavor off-diagonal couplings (which may be vector or axial-vector), and couplings to electroweak gauge bosons through a hypercharge anomaly. In our phenomenological explorations, all three of these varieties of couplings will play a role.

\section{Phenomenology} \label{sec:pheno}

Because the masses of the vector-like heavy quarks are generally far in excess of accessible collider scales, the principal phenomenological signatures of our model will stem from the Goldstone bosons. For our phenomenological analysis, we shall limit ourselves to the case in which the Goldstone bosons are all massless. In principle, one could add arbitrary soft symmetry breaking interactions for the GB's in the second term of Eq.~\eqref{eq:scalar-potential}, which could allow different Goldstone bosons to achieve masses far below their decay constants. Because of the enormous possible parameter space for these terms, plus the potential for such terms to spoil our solution to the strong CP problem, we leave a study of the GB phenomenology with non-negligible mass terms to future work.
In the limit of massless MFV GB's, there are mainly three aspects of phenomenological consequences: cooling stars, other astrophysical searches for massless or light particles, and contributions to the additional radiation degrees of freedom $\Delta N_{\rm eff}$. We discuss all three aspects in turn.  

\subsection{Star Cooling Constraints}
The dominant constraints in our model emerge from limits on the couplings of the GB's to nucleons from supernova (SN)\,1987A~\cite{Carenza:2019pxu}. It is therefore crucial to derive the effective couplings for the GB's to nucleons at a lower energy. Using a general notation to define the axial-vector couplings, one has 
\beqa
\mathcal{L} \supset - \frac{\partial_\mu a}{2\,f_a} \left( c^p_A\, \overline{p}\gamma_\mu \gamma_5 p + c^n_A \,\overline{n} \gamma_\mu \gamma_5 n\right) ~, 
\eeqa
where $p(n)$ denotes a proton(neutron). The dominant contributions to $c^{p, n}_A$ come from the flavor-diagonal coupling in the quark coupling matrices $c^q_A$, with $c^{p, n}_A$ given by 
\beqa
c^{p, n}_A = \sum^3_{i = 1} \sum_{q = u, d, T, B} (c^q_A)_{ii}\, (\Delta q_i)^{p, n} ~.
\eeqa
Here, the matrix coupling $c^q_A$ can be found in~\eqref{eq:GB-couplings} after transforming the GBs into the canonically normalized basis, while the hadronic matrix elements  $(\Delta q_i)^{p, n}$  are given by
\beqa
\langle p | \overline{q}_i \gamma_\mu \gamma_5 q_i | p \rangle = (\Delta q_i)^{p}\, \overline{p}\, \gamma_\mu \gamma_5 p ~, \qquad 
\langle n | \overline{q}_i \gamma_\mu \gamma_5 q_i | n \rangle = (\Delta q_i)^{n}\, \overline{n}\, \gamma_\mu \gamma_5 n ~,   
\eeqa
with $p,\; n$ as the nucleon spinors. At leading order in QCD coupling, the heavy vector-like fermions have zero matrix elements inside a nucleon. At sub-leading order, they do provide threshold effects for the QCD coupling running and indirectly change the relation for the light quark matrix elements evaluated at the UV scale and low-energy scale, \ie\, 2~GeV. Those effects are at the percent level, which we will ignore here. 

The recent lattice QCD results have $(\Delta u)^p = 0.862(17)$ and $(\Delta d)^p = -0.424(16)$ evaluated at 2 GeV~\cite{Alexandrou:2019brg}. Using the isospin symmetry, we have the couplings of the two GB's $a_d$ and $a_u$ to nucleons are 
\beqa
\mathcal{L} \supset - \frac{\partial_\mu a_d}{2f_{a_d}}\, \left( -0.424\,\overline{p}\gamma_\mu \gamma_5 p + 0.862\, \overline{n} \gamma_\mu \gamma_5 n \right) 
- \frac{\partial_\mu a_u}{2f_{a_u}}\, \left( 0.862\,\overline{p}\gamma_\mu \gamma_5 p -0.424\, \overline{n} \gamma_\mu \gamma_5 n \right) \,. 
\eeqa
Requiring the GB's emission luminosity smaller than the neutrino luminosity for SN\,1987A and using Eq.~(3.3) in Ref.~\cite{Carenza:2019pxu}, the constraints on the independent GB decay constants are
\beqa \label{eq:fau-fad-SN-constraints}
f_{a_d} \gtrsim 8.4 \times 10^8\, \mbox{GeV} \,, \qquad \qquad f_{a_u} \gtrsim 6.8 \times 10^8\, \mbox{GeV} ~.
\eeqa
Using the quark mass ratios extracted from Eq.~(\ref{eq:quark-masses}) and the relations of $f_{a_d} \equiv f_d\times m_d/m_b$ and $f_{a_u} \equiv f_u\times m_u/m_t$, the independent constraints on the VEV's are 
\beqa \label{eq:fu-fd-SN-constraints}
f_d \gtrsim 7.7 \times 10^{11}\,\mbox{GeV} \,, \qquad \qquad f_u \gtrsim 7.1 \times 10^{13}\,\mbox{GeV}\,.
\eeqa

In reality, it is impossible to isolate the effects of a single GB in our model from those of the 25 others that contribute to stellar energy loss. However, generalizing the constraint of Ref.~\cite{Carenza:2019pxu} to the case of multiple GBs is straightforward: assuming that the GBs have negligible interactions among one another (which is reasonable given that the temperature in the star is much lower than all phenomenologically feasible GB decay constants here), then to good approximation each GB's energy loss will simply add to the total. Depending on the relative values of $f_d$ and $f_u$, then, we can easily combine the energy loss contributions of both $a_d$ and $a_u$ using a trivial generalization of Eq. (3.3) of Ref.~\cite{Carenza:2019pxu}, in order to derive a somewhat more realistic constraint on $f_u$ given $f_d/f_u$. In practice, the  limit on $f_u$ given in Eq.~(\ref{eq:fu-fd-SN-constraints}) is extremely accurate for $f_d/f_u \gtrsim O(1)\times m_b/m_t$, and experiences only $O(1)$ modifications if we bring $f_d$ down to its minimum value $f_d/f_u \sim m_b/m_t$.

Of course, even if we include the combined effect of $a_d$ and $a_u$ on stellar energy loss, the effects of the remaining 24 GBs should also in principle be accounted for. However, in practice this is not significant: because the decay constants of the other GBs are hierarchically larger than those of $a_u$ and $a_d$, we find numerically that the other GBs only contribute to the total GB-mediated stellar energy loss at the sub-percent level.

Finally, we may note that the nucleon coupling constraints discussed in this section assume that the GBs are all too weakly coupled to be trapped within the supernova after being emitted. If instead the decay constants $f_{a_u}$ and $f_{a_d}$ were \emph{small} enough to ensure that $a_u$ and $a_d$ would not escape the supernova after emission, naively we might assume that all constraints on $f_u$ and $f_d$ from the supernova observations would be avoided, as can occur in scenarios with a single GB \cite{Lee:2018lcj}. Because the other GBs in our model are hierarchically more weakly coupled than $a_u$ and $a_d$ to matter, however, avoiding the SN 1987A constraint in this manner is considerably more difficult in our construction than in the scenario with a single GB: the region of parameter space with $a_u$ and $a_d$ trapped in the supernova will still radiate a significant amount of energy through emission of other, more weakly-coupled GBs. Moreover, if $f_u$ and $f_d$ are allowed to decrease enough to trap $a_u$ and $a_d$, constraints from flavor observables, most notably the branching ratio $\mbox{Br}(K\rightarrow \pi + X)$ \cite{BNL-E949:2009dza,KOTO:2018dsc}, where $X$ is some invisible particle(s) (in our case, a flavor-changing GB), will generally disallow this region of parameter space. In practice, therefore, the limits given in Eqs.~(\ref{eq:fau-fad-SN-constraints}) and (\ref{eq:fu-fd-SN-constraints}) represent the only region of parameter space which can evade the supernova constraints and remain otherwise phenomenologically viable.

\subsection{Other Astrophysical Constraints}

In addition to the strong constraints on GB-matter couplings from observations of SN 1987A, the GB-photon couplings can be constrained from independent astrophysical observations. At present, the best limits on GB couplings to photons come from observations of relative populations of Asymptotic Giant Branch and Horizontal Branch stars in globular clusters, which are sensitive to the energy loss undergone when stars emit GBs~\cite{Dolan:2022kul}. As in the case of the nucleon couplings, we find numerically that the dominant contributions to these constraints will invariably stem from the GBs $a_u$ and $a_d$, however, in contrast to the nucleon couplings, these observations will be dependent on the hypercharge assignments $Q_B$ and $Q_T$ for the vector-like quarks $B$ and $T$, respectively. Assuming that only $a_u$ and $a_d$ contribute significantly to stellar energy loss, the constraint extracted from Ref.~\cite{Dolan:2022kul} becomes
\beqa
\bigg\{ \frac{1}{f_{a_u}^2} \bigg[\left(\frac{2}{3}\right)^2-Q_T^2\bigg] + \frac{1}{f_{a_d}^2} \bigg[\left(\frac{1}{3}\right)^2-Q_B^2\bigg] \bigg\} \frac{9\, \alpha_{\rm em}^2}{2 \pi^2} \leq 2.2 \times 10^{-21} \; \textrm{GeV}^{-2}.
\eeqa
It is easiest to get a feel for these constraints by considering the individual contributions of $a_d$ and $a_u$ for various selections of the $Q_B$ and $Q_T$, respectively. Assuming that the electromagnetic charge of the vector-like quarks only differs from that of their corresponding SM quarks by $\pm 1$, quantum of electric charge, we have 
\begin{align}
\renewcommand{\arraystretch}{1.3}
    \begin{matrix}
        f_{a_u}|_{Q_T=\frac{5}{3}} \gtrsim 2.4 \times 10^8 \; \textrm{GeV}, & \qquad  f_{a_u}|_{Q_T=-\frac{1}{3}} \gtrsim 3.5 \times 10^8 \; \textrm{GeV},\\
        f_{a_d}|_{Q_B=\frac{2}{3}} \gtrsim 3.5 \times 10^7 \; \textrm{GeV}, & \qquad f_{a_d}|_{Q_B=-\frac{4}{3}} \gtrsim 1.8 \times 10^7 \; \textrm{GeV}.
    \end{matrix}
\end{align}
In terms of scales $f_u$ and $f_d$, these constraints become
\beqa
\renewcommand{\arraystretch}{1.3}
    \begin{matrix}
        f_u|_{Q_T=\frac{5}{3}} \gtrsim 3.2 \times 10^{13} \; \textrm{GeV}, &\qquad  f_u|_{Q_T=-\frac{1}{3}} \gtrsim 4.7 \times 10^{13} \; \textrm{GeV},\\
        f_d|_{Q_B=\frac{2}{3}} \gtrsim 3.7 \times 10^{10} \; \textrm{GeV}, &\qquad f_d|_{Q_B=-\frac{4}{3}} \gtrsim 1.8 \times 10^{10} \; \textrm{GeV}.
    \end{matrix}
\eeqa
The constraints vanish for the special choice with $|Q_T| = 2/3$ and $|Q_B| = 1/3$.

\subsection{Additional Radiation Degrees of Freedom: $\Delta N_{\rm eff}$}\label{sec:delta-n-eff}
Apart from modern astrophysical constraints, we also find that phenomenological consequences of this model affect cosmological probes. The strongest constraints for the current model come from relic abundances of hadrons formed from heavy vector-like quarks, since the lightest up-like and down-like vector-like quarks are stable, however as we shall discuss at the end of this Section, the model can be easily extended to destabilize these fermions and evade these constraints. A more robust cosmological probe comes from the additional radiation degrees of freedom stemming from the Goldstone bosons:
if any of the GBs are thermally produced in the early universe, they will constitute additional dark radiation, to which the cosmic microwave background (CMB) is sensitive---this sensitivity is usually quantified with the parameter $\Delta N_{\rm eff}$, the ratio of dark radiation energy density to that of a single relativistic SM neutrino species at the time of photon decoupling. As the precision of the constraint on $\Delta N_{\rm eff}$ improves~\cite{Abazajian:2019eic}, it is feasible that the cosmological signature of this model will be observable even for values of $f_u$ that are far in excess of what can be constrained in the conventional astrophysical searches discussed in previous sections.
As such, it is useful to explore how $\Delta N_{\rm eff}$ may be modified in our model.

To estimate the contribution of a single Goldstone boson $a$ to $\Delta N_{\rm eff}$, we can simply use the relation~\cite{Green:2021hjh}
\beqa\label{eq:delta-n-eff-formula}
\Delta N_{\rm eff} = \frac{4}{7} \bigg( \frac{11 \pi^4}{90\,\zeta(3)} g_{*s,\infty}\,Y_{a, \infty}\bigg)^{\frac{4}{3}}~,  \qquad\,\mbox{(for a single GB)} ~,
\eeqa
where $Y_{a, \infty}$ is the relic yield of $a$ (the number density divided by the entropy density of the universe) after recombination, and $g_{*s, \infty}=43/11$ is the effective number of relativistic entropy degrees of freedom of the universe after recombination. Then, we only need to find $Y_{a, \infty}$ for each GB in our model, which can be accomplished by solving the Boltzmann equations for all 26 Goldstone bosons. We shall sketch the conceptual outlines of this computation here, and direct the reader to Appendix~\ref{sec:thermal-GB} for a more detailed discussion of the assumptions and methods we have employed. For our purposes, we find that the relic yields of the Goldstone bosons can be well-modeled by considering 26 GB's as the solutions to separate uncoupled Boltzmann equations, of the form
\beqa\label{eq:boltzmann-eq}
\frac{d n_a}{d t} + 3\,H(T)\,n_a = \Gamma_a (T)\, \Bigl(n_a^{\rm eq}(T) - n_a\Bigr) ~,
\eeqa
where $H$ is the Hubble parameter, $n_a$ is the number density of some GB $a$, $n_a^{\rm eq}$ is the number density of this GB in thermal equilibrium, and $\Gamma_a (T)$ is a production rate for the GB $a$.

The production rate $\Gamma_a(T)$ is determined by processes in three broad categories: $2 \rightarrow 2$ scattering of SM quarks, decays of the heavier vector-like quarks into lighter vector-like quarks plus a GB, and $2 \rightarrow 2$ scattering of same-flavor vector-like quarks.~\footnote{For simplicity, we shall ignore gauge boson couplings to the Goldstone bosons in this section---because the gluons lack any coupling to the Goldstones by construction, and the photon-Goldstone couplings are highly model-dependent and can be easily made to vanish. We find this assumption reasonable for the estimations we make here.}

Given the radical difference between the mass scales of the SM fermions and those of the heavy vector-like quarks, our computations of the rates of these different classes of processes will make different approximations in order to maximize accuracy while keeping our results as simple as possible---the specifics of these approximations and their motivations will be discussed in Appendix \ref{sec:thermal-GB}.

\begin{figure}[th!]
	\includegraphics[width=1.0\textwidth]{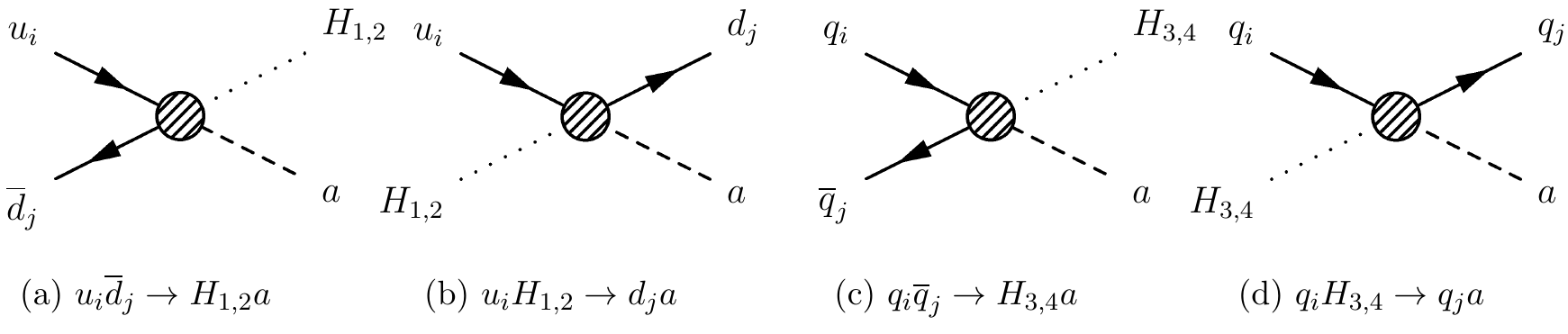}
    \caption{The four processes which dominate the production rate of a Goldstone Boson $a$ from its couplings to SM quarks. Here, the indices $i$ and $j$ represent flavor indices, and $q=u,d$ can denote an up-like or down-like quark. $H^T = (H_1 + i H_2, H_3 + i H_4)/\sqrt{2}$ denotes the SM Higgs field before electroweak symmetry breaking.}
    \label{fig1}
\end{figure}

For the $2 \rightarrow 2$ scatterings featuring SM quarks in Figure~\ref{fig1}, our analysis closely follows that of Ref.~\cite{Baumann:2016wac}, which considered a single Goldstone boson with generic derivative couplings to quarks and/or leptons. We find that to excellent approximation, the production rate for the Goldstone boson $a$ from $2 \rightarrow 2$ scattering of SM quarks is given by
\beqa \label{eq:SM-quarks-rate}
    \Gamma_a^{\rm SM}(T) = \frac{T^3}{f_u^2} \bigg[ \bigg(\frac{7}{8} \bigg)^2\, \frac{9\,\zeta(3)}{16 \pi^3} + \frac{3\pi}{1152 \,\zeta(3)}\bigg] \,\frac{2\,m_t^2}{v^2} \,\bigg( \frac{m_b^2}{m_t^2} G^d_{\rm SM}+ G^u_{\rm SM} \bigg)\,, \\ [5pt]
    G^{u,d}_{\rm SM} \equiv \mbox{Tr} [(g^{u,d}_L \, R_{u,d} - R_{u,d} \,g^{u,d}_R)(R_{u,d} \,g^{u,d}_L - g^{u,d}_R \,R_{u,d})]\,, \nonumber
\eeqa
where $g^{u,d}_{L,R}$ are the coupling matrices defined in Eq.~(\ref{eq:generic-SM-couplings}), in Appendix \ref{sec:GB}.

\begin{figure}[th!]
	\includegraphics[width=1.0\textwidth]{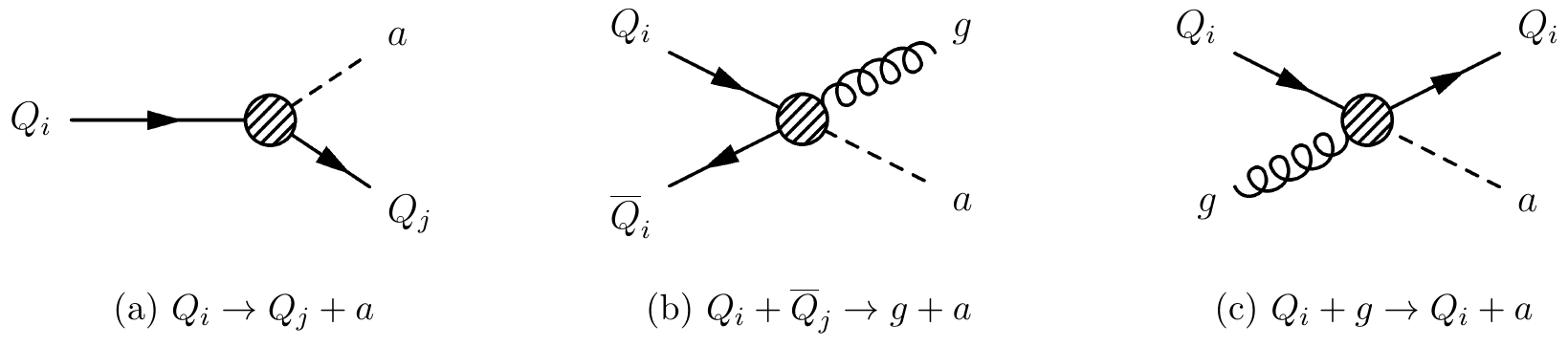}
    \caption{The three processes which dominate the production rate of a Goldstone boson $a$ from its couplings to vector-like heavy quarks: (a) decay of heavier vector-like quarks into lighter ones,  (b) quark-antiquark annihilation, and (c) Compton-like scattering.} \label{fig2}
\end{figure}

Meanwhile, the vector-like quark decays and $2 \rightarrow 2$ scattering processes are depicted in Figure \ref{fig2}. All of these processes have immediate parallels in similar processes for SM quarks at much lower temperatures in, \eg, \cite{Masso:2002np,Baumann:2016wac,Green:2021hjh}---they are the production of a Goldstone boson from the decay of a heavier vector-like quark to a lighter (still vector-like) one, the production in association with a gluon from quark-antiquark annihilation, and the production from the scattering of a gluon off of a quark.
A notable characteristic of these rates is that, just as observed in \cite{Masso:2002np,Baumann:2016wac,Green:2021hjh} and in contrast to the production rate from the SM quark scattering, the rates associated with vector-like quark interactions will scale more slowly with temperature in certain regions than the Hubble rate, allowing for a ``freeze-in'' of a previously decoupled Goldstone bosons at temperatures near the masses of the vector-like quarks. Specific formulas for the rates emerging from the processes in Figure \ref{fig2} are lengthy, but given in full in Eqs.~(\ref{eq:1to2-rate-VLQs}-\ref{eq:2to2-gamma-functions}) in Appendix \ref{sec:thermal-GB}.

\begin{figure}[th!]
    \centerline{
    \includegraphics[width=0.58\textwidth]{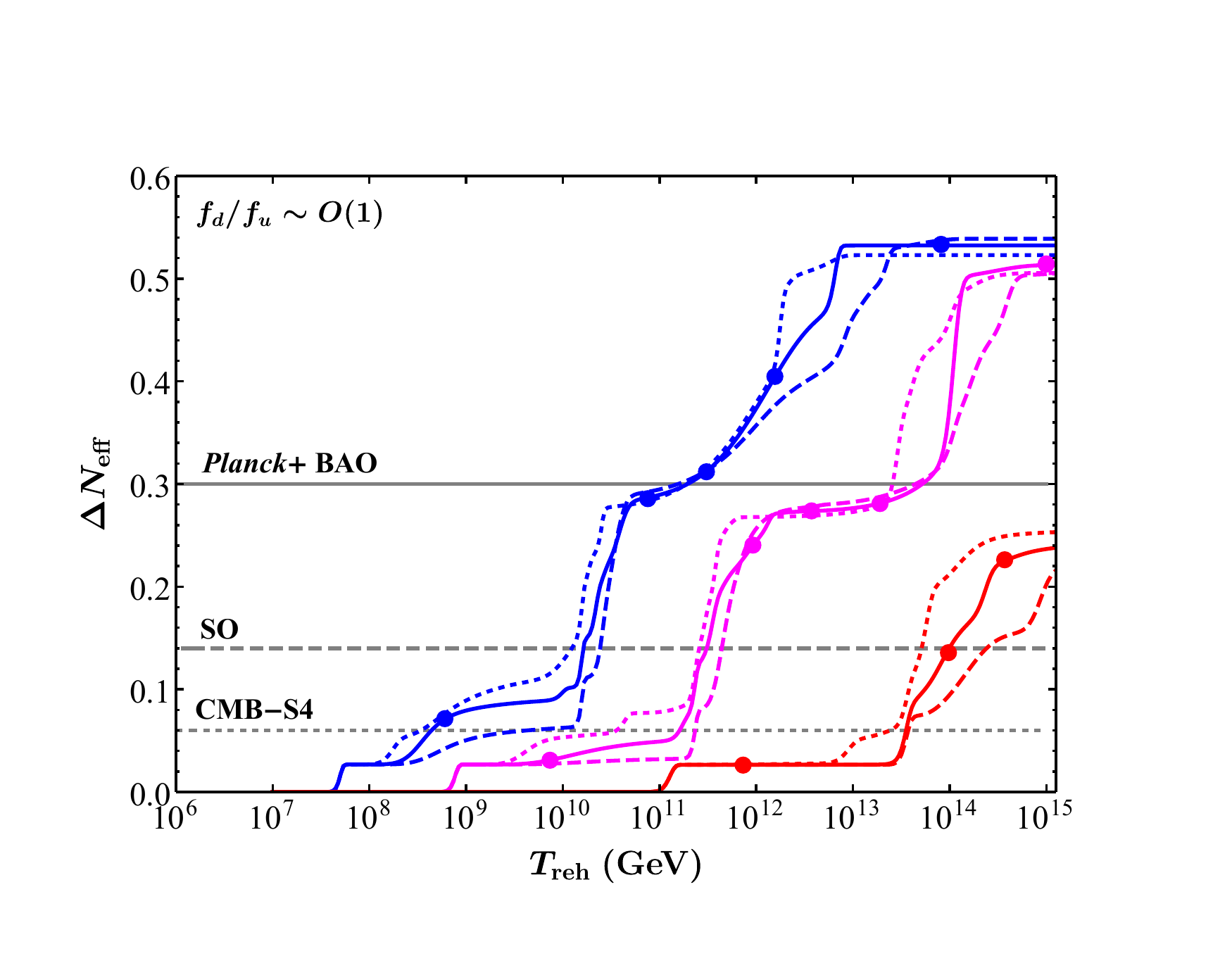}
    \hspace{-1.5cm}
    \includegraphics[width=0.58\textwidth]{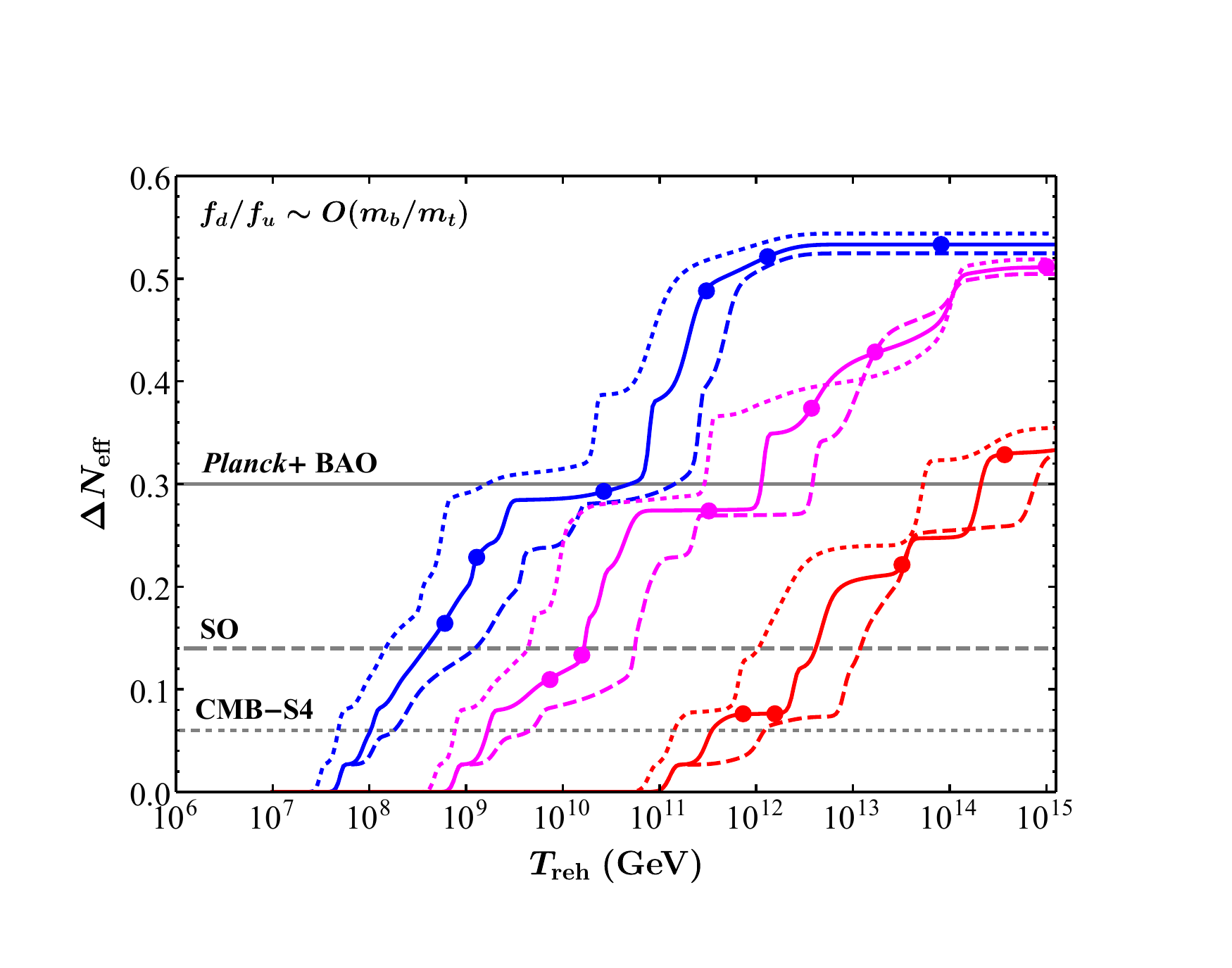}}
    \caption{ 
    Left panel: the ``{\it flavor stairway}" plot or the predicted $\Delta N_{\rm eff}$ for the model as a function of the reheating temperature $T_{\rm reh}$, assuming that $f_u= 8\times 10^{13} \; \textrm{GeV}$ (blue), $10^{15} \; \textrm{GeV}$ (magenta), and $10^{17} \; \textrm{GeV}$ (red), and $f_d=f_u$ (solid), $f_d = 3 f_u$ (dashed), and $f_d = 0.3 f_u$ (dotted). The Yukawa couplings for the $T$ and $B$ quarks are both taken to be unity. On the solid lines, the dots indicate the points at which  $T_{\rm reh}$ is equal to a heavy vector-like quark mass. Note that when $f_d=f_u$, the masses of the two heaviest vector-like quarks are degenerate, so we may only expect a maximum of 5 dots to appear on each solid line. The gray solid line indicates the $95\%$-CL upper limit of $\Delta N_{\rm eff}$ from \emph{Planck} and BAO \cite{Planck:2018vyg}, while projected $95\%$-CL limits are depicted from the Simons Observatory \cite{SimonsObservatory:2018koc} (dashed) and CMB-S4 \cite{Abazajian:2019eic} (dotted). Right panel: the same as the left one, but with $f_d = (m_b/m_t) f_u$ (solid), $f_d = 3 (m_b/m_t) f_u$ (dashed), and $f_d = (m_b/m_t) f_u$ (dotted).  }
    \label{fig3}
\end{figure}

In Figure \ref{fig3}, we depict the results of our $\Delta N_{\rm eff}$ calculations for various values of $f_u$ and $f_d/f_u$ as a function of the reheating temperature $T_{\rm reh}$. To obtain a better sense of the significance of the heavy vector-like quarks in the results, we mark the points along the solid lines to denote where $T_{\rm reh}$ is equal to a heavy vector-like quark mass with large dots. The strong correlation between the mass spectrum of the heavy vector-like quarks and the predicted $\Delta N_{\rm eff}$ in Figure \ref{fig3} illustrates a key characteristic of the thermal production of the Goldstone bosons: in regions which satisfy constraints from \emph{Planck}, $\Delta N_{\rm eff}$ is dominated by the interactions with the heavy vector-like quarks---Goldstone bosons begin to couple to the thermal bath at temperatures as much as an order of magnitude below the mass of the lightest vector-like quark, while in the absence of vector-like quarks we find that GB thermal production only occurs at reheating temperatures several orders of magnitude higher. In principal, we could increase the temperatures at which the vector-like quark processes couple the GB's to the thermal bath by increasing the vector-like quark Yukawa couplings, which in Figure \ref{fig3} we have assumed to be equal to unity. However, even if both Yukawa couplings were increased to the perturbativity limit $\sim 4 \pi$, the GB production processes from SM quark interactions will still freeze out at far higher temperatures than the ones from vector-like quarks. Because the Goldstone boson production rates are dominated by the processes involving vector-like quarks, we can further see that the predicted $\Delta N_{\rm eff}$ is largely dictated by these heavy quarks' mass spectrum: if the reheating temperature is high enough to produce meaningful number densities of 3 species of vector-like quarks, this is generally sufficient to yield a $\Delta N_{\rm eff}$ approximately at the current \emph{Planck} constraint \cite{Planck:2018vyg}, while meaningfully producing one or two of the vector-like quark species can easily produce a $\Delta N_{\rm eff}$ value that would be observable by near-future probes \cite{SimonsObservatory:2018koc,Abazajian:2019eic}.

It is also interesting to note that Goldstone bosons have a tendency to couple to the SM thermal bath in ``clumps'' at various temperatures, with the clumping behavior dictated by the SM flavor structure and the ratio $f_d/f_u$. This is particularly apparent in Figure \ref{fig3} (which we call the ``flavor stairway" plot), where the lightest vector-like quark (the up-quark's vector-like counterpart) is much less massive than all the other vector-like quarks. So, those Goldstone bosons most strongly coupled to the first generation up-like quarks will achieve a significant relic abundance, and contribute significantly to $\Delta N_{\rm eff}$, at much lower reheating temperatures than the other Goldstone bosons do. This in turn leads to a distinctive plateau in Figure~\ref{fig3}. For $f_u = 8 \times 10^{13} \; \textrm{GeV}$ and $f_{a_u} = 6\times 10^8$~GeV, this plateau occurs for $T_{\rm reh} \sim 10^{9-10} \; \textrm{GeV}$, at which point some Goldstone bosons are contributing significantly to $\Delta N_{\rm eff}$ while others
remain uncoupled from the SM bath. The plateau ends as a large number of new Goldstone bosons achieve significant abundances over a small increase in the reheating temperature for $T_{\rm reh} \gtrsim 10^{10} \; \textrm{GeV}$. Similar plateaus, varying depending on the mass hierarchies of the various vector-like quarks, are apparent elsewhere in Figure~\ref{fig3} as well. This suggests that points in parameter space of this model will have certain ``natural'' values of $\Delta N_{\rm eff}$, which might be achieved over a large range of reheating
temperatures, and other ``finely-tuned'' $\Delta N_{\rm eff}$ values that would require quite precise values of the reheating temperature for the model to achieve. This is particularly notable since many of these natural values are potentially discoverable by near-term future cosmological measurements even for values of $f_u$ and $f_d$ that are far in excess of any constraints from present-day astrophysical constraints, such as stellar cooling and SN 1987A.

Finally, as noted at the beginning of this Section, a reader may also be concerned about the fact that $\Delta N_{\rm eff}$ is only appreciably modified if the reheating temperature is at least comparable to the masses of one or more of the vector-like quarks. In the theory as written, the lightest up-like and down-like vector-like quarks are stable, so we would then anticipate that they would achieve an appreciable relic abundance. This could present significant cosmological problems, most notably that the vector-like quarks' high masses (and therefore low annihilation cross sections) will generically severely overproduce them.
However, these same large masses mean that an extremely feeble interaction to mediate decay into SM quarks can be used to ensure that the vector-like quarks decay rapidly, long before BBN. We could achieve this, for example, by introducing extremely small off-diagonal mass terms in the mass matrices of Eq.~(\ref{eq:mass-matrix}), modifying the up-type quark mass matrix to the form
\beqa
\renewcommand{\arraystretch}{1.3}
    \mathcal{M}^u = 
    \begin{pmatrix}
    \dfrac{y_u\,v}{\sqrt{2}}\, \dfrac{\langle \Sigma_u \rangle}{\Lambda} & \dfrac{y_1 v}{\sqrt{2}} \dfrac{\mathcal{M}_1}{\Lambda} \\
    y_2 \mathcal{M}_2 &  \eta_u \, \langle \Sigma_u \rangle^* 
    \end{pmatrix} ~,
\eeqa
where $\mathcal{M}_{1,2}$ are arbitrary mass matrices (that might emerge from either explicit symmetry breaking terms or the introduction of an additional scalar, for example in a representation $(3,1,3)_{+1}$ of $SU(3)_{q_L} \times SU(3)_{d_R} \times SU(3)_{u_R} \times U(1)_u$, with a VEV far below $\langle \Sigma_{u,d} \rangle$). We also will assume that the large mass scale $\Lambda$ for the effective operators remains $\sim f_u$, as we have done earlier when 
motivated by the need to reproduce the top quark mass. Assuming that we parameterize these mass terms such that $\mathcal{M}_{1,2} \sim \langle \Sigma_u \rangle$ and $y_{1,2} \ll 1$,  a heavy quark decay width into an SM quark and an electroweak boson is suppressed as $O(y_1^{2}, y_2^{2}, y_{1} y_{2})$. We find that a vector-like quark of mass $O(10^8 \; \textrm{GeV})$, the smallest mass consistent with astrophysical constraints, achieves a lifetime of $\sim 10^{-4} \; \textrm{s}$ for $y_{1,2} \gtrsim 10^{-9}$ and allowes them to decay well before BBN. Meanwhile, the tree-level correction to the strong CP phase from these terms will be approximately
\beqa
    -\frac{y_1 y_2}{y_u \eta_u} \mbox{Im} \mbox{Tr}\bigl[\mathcal{M}_1 \langle \Sigma_u^* \rangle^{-1} \mathcal{M}_2 \langle \Sigma_u \rangle^{-1}\bigr] \sim O(y_1 y_2)~,
\eeqa
so a sufficiently rapid decay of the vector-like quarks can be achieved without correcting the strong CP phase by more than $\sim O(10^{-18})$.~\footnote{If we were to take the idea of a scalar VEV generating this  off-diagonal mass seriously, we would also have to include insertions of these VEV's into higher-dimensional operator corrections to the blocks in our mass matrices. However, since these operators would only appear at the $O(y_1^2,y_2^2,y_1 y_2)$ level, we would anticipate that their corrections to the strong CP phase would be even smaller than the tree-level contribution.} While this method of allowing the vector-like quarks to decay is somewhat vague and cursory, it does demonstrate that avoiding thermal overproduction of the new fermions even with reheating temperatures comparable to heavy quark masses is entirely plausible with a minimal extension.

\section{Discussion and Conclusions}\label{sec:conclusion}

In this work, we have presented a straightforward realization of a basic class of strong CP models that has not thus far gained significant attention: models of spontaneous CP violation in which a continuous flavor symmetry enforces cancellation between complex determinants in the SM quark mass matrices and those appearing in the mass matrices of new QCD-charged fermions. By selecting the global flavor symmetry to be that of minimal flavor violation in the quark sector, we have ensured that the only physical CP-violating phase in the model is the weak CKM phase, thus guaranteeing that even under higher-order corrections, $\overline{\theta} \lesssim O(10^{-13}-10^{-24})$, depending on model parameter values. Our large global symmetry (not including any symmetries in the lepton sector) has given rise to no fewer than 26 Goldstone bosons with flavorful couplings. We have studied the phenomenology of the case in which these Goldstone bosons are massless, and have found harsh constraints on the CP and flavor-symmetry breaking scale from limits on axion-like-particle couplings to nucleons from SN 1987A energy loss. Perhaps more intriguingly, we have also found that the large number of Goldstone bosons can yield a contribution to $\Delta N_{\rm eff}$, which are potentially observable at near-future CMB experiments even if the Goldstone bosons themselves are far too weakly coupled to be detectable via astrophysical processes. Furthermore, the spectra of the heavy vector-like quarks, which are ultimately controlled by SM quark mass hierarchies, result in the Goldstone bosons' contributions to $\Delta N_{\rm eff}$ forming a distinctive ``flavor stairway'' as a function of the cosmic reheating temperature---plateaus of natural values over a wide range of reheating temperatures, punctuated by regions where $\Delta N_{\rm eff}$ rises rapidly.

There are several directions through which the specific construction we have studied here might be explored further. First, the principal remaining challenge of this model is to obtain the scalar VEV structures consistent with the SM flavor structure, a highly non-trivial task that may require soft symmetry-breaking terms in the potential and/or additional scalars (see Refs.~\cite{Alonso:2011yg,Espinosa:2012uu,Fong:2013dnk} for some attempts). Notably, to prevent unacceptable corrections to the strong CP phase (and maintain the validity of the MFV paradigm), any new scalars must have VEVs either hierarchically smaller than the VEV's of the scalars $\Sigma_{u,d}$ or somehow decoupled from the QCD-charged fermions in the model (for example due to their group representations), such that they do not affect fermion mass matrices significantly. Similarly, the soft symmetry-breaking terms in the potential must also be limited for the same reason. Because these questions are related to the physics at the very high flavor-symmetry breaking scale, however, it is improbable that such difficulties will have any bearing on the low-energy phenomenology of the model as long as such solutions exist. 

A more phenomenologically relevant direction of work would lie in considering the case in which some or all of the Goldstone bosons of the model have some non-trivial masses. Depending on the mass spectrum generated by, \eg, soft symmetry breaking terms or a feeble gauging of all or part of the global flavor symmetry, leading experimental constraints on the scale of the CP and flavor-symmetry breaking scalar VEV's can be considerably altered. For example, Goldstone boson masses $\gtrsim 30 \; \textrm{MeV}$ will generally have much weaker constraints from supernova energy loss than the case in the main text. Furthermore, depending on the spectrum of the Goldstone bosons, the dominant phenomenological constraints on the model parameter space might stem from a variety of precision flavor measurements, similar to the studies on flavorful axions~\cite{Arias-Aragon:2017eww,Green:2021hjh,Baumann:2016wac,delaVega:2021ugs,Bauer:2021mvw}. The cosmological implications in the event that the Goldstone bosons have significant masses will also depend strongly on these masses' magnitudes and relative scales---sufficiently long-lived Goldstone bosons may still have an effect on $\Delta N_{\rm eff}$, but may also affect measurements of the sum of neutrino masses or large scale structure formation \cite{Dvorkin:2022jyg,Xu:2021rwg}.

Before concluding, it is interesting to broaden our discussion again and consider other possibilities in which the SM quarks' contribution to the strong CP phase is cancelled by that of some new heavy QCD-charged fermions due to a continuous flavor symmetry. Enumerating all scenarios in which Eq.~(\ref{eq:general-cond}) is satisfied is, of course, a highly non-trivial problem and well beyond the scope of this paper, given that such a cancellation might be theoretically realized by an arbitrary number of heavy QCD-charged fermions with arbitrary representations under QCD and the flavor group (and, indeed, arbitrary flavor groups), so long as Eq.~(\ref{eq:general-cond}) was ultimately satisfied. However, we can briefly comment on how some aspects of our MFV-inspired construction might apply more generally to other similar models.

First, we note that our selection of a very large flavor group [specifically $SU(3)^3 \times U(1)^2$] is ultimately what insulates the model in this work from unacceptably large corrections to the strong CP phase. Taking smaller flavor groups, as was done in Ref.~\cite{Vecchi:2014hpa} to satisfy Eq.~(\ref{eq:general-cond}) \emph{without} introducing new heavy vector-like quarks, tend toward introducing larger corrections to the strong CP phase, since additional physical CP-violating phases enter the theory. Given the harsh constraints on the strong CP phase, we can therefore conjecture that any phenomenologically realistic model in this class will involve a large continuous flavor symmetry, and therefore a large number of associated Goldstone bosons. If the Goldstone bosons are massless (or very nearly so), as in our phenomenological analyses in Section \ref{sec:pheno}, we can anticipate marked similarities between the results of that section and what we might anticipate in other models of this class. Astrophysical and collider constraints on these Goldstone bosons would then stem from their couplings to SM quarks, which would ultimately be determined by the SM flavor structure, just as in the specific model presented in this work. 

Similarly, the large number of Goldstone bosons in any viable model suggests that the cosmological signatures of constructions in this class will be analogous to the results in this work, in particular the contribution to $\Delta N_{\rm eff}$ in Figure \ref{fig3}. Since suppressing quantum corrections to the strong CP phase likely requires a large continuous flavor symmetry corresponding to many Goldstone bosons, we should generically anticipate that for any such construction $\Delta N_{\rm eff}$ contributions will be large enough to be potentially within the reach of near-future cosmology probes even for exceptionally high flavor-symmetry breaking scales, provided the reheating temperature is sufficiently high. While the quantitative aspects of our results in Figure~\ref{fig3} are limited to only the specific model we consider in the main text, we might also use them to inform our expectations of the $\Delta N_{\rm eff}$ contribution in models within this modified Nelson-Barr paradigm. In particular, it is clear from our results in Figure \ref{fig3} that the heavy QCD-charged fermions play a decisive role in thermally coupling the Goldstone bosons to the photon bath if the reheating temperature is high enough to produce these fermions in abundance. Intuitively, we can see this occurs because the derivative coupling of a Goldstone boson to a fermion is related (by the equations of motion) to the fermion mass, and the new heavy fermions' masses are far greater than those of the SM quarks. We see from Figure \ref{fig3} that the flavor Goldstone bosons will virtually always achieve equilibrium with the SM bath as long as the heavy fermion to which it is dominantly coupled is produced in the thermal bath. Since we are specifically considering constructions in which at least one of these heavy fermions must exist to cancel the strong CP phase of the SM quarks, this suggests that at least some variant of the ``flavor stairway'', in which $\Delta N_{\rm eff}$ exhibits ``plateaus'' of natural values over a wide range of reheating temperatures. A flavor stairway quite similar to the one presented in Figure \ref{fig3} will emerge if there exists a mass hierarchy between different new heavy fermions, as additional Goldstone bosons couple to the thermal bath once the reheating temperature becomes large enough to produce a given heavy fermion. An analogous flavor stairway can even occur in models with a single heavy fermion, or those without a significant hierarchy of heavy fermion masses. In that case, the flavor stairway will emerge from a hierarchy in the couplings of different Goldstone boson species to the heavy fermion(s)---those Goldstone bosons that are more weakly coupled to the heavy fermions will only be thermalized with the SM bath at significantly higher reheating temperatures than those which are more strongly coupled. The hierarchical SM flavor structure renders it highly likely for hierarchies to appear in either the Goldstone boson couplings or the spectrum of heavy fermions; therefore, while its existence is difficult, if not impossible, to avoid, we should note that the specific form of the flavor stairway will depend heavily on the particular flavor group and the representations of the new heavy fermions under it.

In short, the paradigm we have presented here, in which a continuous flavor symmetry enforces a cancellation between complex phases in the SM quark mass matrix determinants and those of new heavy vector-like quarks, presents an interesting space of models for which the strong CP problem might be resolved, as shown in the sample model discussed in this work. Further exploration of the space of models in this class which satisfy experimental constraints on both the strong CP phase and the Goldstone bosons is a highly non-trivial, but intriguing, problem which merits further exploration.

\vspace{1cm}
\subsubsection*{Acknowledgments}
The work is supported by the U.S.~Department of Energy under the contract DE-SC-0017647.

\appendix
\section{Operators Generating Strong Phase from Weak Phase}
\label{sec:more-operators}

Here we present a detailed exploration of all possible higher-dimensional operators that might spoil the cancellation of the complex phases in the SM quark matrix with that of the new vector-like heavy quarks. To illustrate that these operators will not contribute significantly to the strong CP phase, we can leverage the freedom afforded by our global symmetries to choose a suggestive basis for $\langle \Sigma_u \rangle$ and $\langle \Sigma_d \rangle$. Specifically, we can use a $SU(3)_{q_L} \times SU(3)_{d_R} \times SU(3)_{u_R} \times U(1)_u \times U(1)_d$ transformation to write these VEV's as
\begin{align}\label{eq:SigmaDefs}
\begin{matrix}
    \langle \Sigma_u \rangle = f_u \,V_{\rm CKM}^\dagger\,\cdot\,\mbox{diag}\left( \dfrac{m_u}{m_t}, \dfrac{m_c}{m_t}, 1 \right), & \quad \langle \Sigma_d \rangle = f_d \,\mbox{diag} \left( \dfrac{m_d}{m_b}, \dfrac{m_s}{m_b}, 1 \right),
\end{matrix}
\end{align}
where $V_{\rm CKM}$ is the CKM matrix and $m_{u,c,t,d,s,b}$ are the masses of the SM quarks. Note that these definitions of $\langle \Sigma_{u,d} \rangle$ guarantee the correct mass eigenvalues for the SM quarks using the mass matrices of Eq.~(\ref{eq:mass-matrix}), as long as $y_u = (\sqrt{2}\,m_t \,\Lambda)/(v\,f_u)$ and $f_d/f_u = (y_u m_b)/(y_d m_t)$ in that expression. In this basis, we can see that the \emph{only} source of CP violation in the theory lies in the CKM phase or the weak phase---this is ultimately a consequence of our selection of the MFV group as our global symmetry and the assumption that only the scalars $\Sigma_{u,d}$ contribute to the mass matrices of the SM and vector-like heavy quarks, at least to  a very good approximation. An actual computation of the radiative corrections to $\overline{\theta}$ in this model would be extremely involved, since leading corrections only enter at high loop order. Furthermore, we have at best a limited knowledge of the scalar sector of the model, since we are agnostic about the form of the potential generating the VEV's of $\Sigma_{u,d}$. However, we might estimate the degree of suppression of these corrections by noting that all such corrections must feature insertions of the $\Sigma_{u,d}$ VEV's to mediate CP violation. We can therefore estimate the suppression of our corrections via ``spurion'' insertions of $\langle \Sigma_{u,d} \rangle$, scaled by some UV cutoff. Our task now remains to classify the operators that might communicate the non-trivial CKM phase to the strong CP phase.

Assuming our setup remains valid up to a UV cutoff of $\sim f_u$ (or $4\pi f_u$), (the characteristic scale of both the Goldstone boson interactions and the dimension-5 operator granting the SM quarks mass~\footnote{A reader may be concerned that we have a second scale, $f_d$, which in principle may be much lower than $f_u$. Perturbativity suggests that we can't insert a spurion $\langle \Sigma_u \rangle/f_d$ in this case, but we might insert spurions of $\langle \Sigma_d \rangle/f_d$ here. Numerically this will have the same effect as rescaling $f_d$ in our analysis, so we don't consider it explicitly here.}), inspection of the representations of our scalars $\Sigma_u$ and $\Sigma_d$ indicates that the leading corrections to the complex phase of the SM up-like quark mass matrix (with analogous results for the heavy up-like vector-like quarks, and the down-like SM and vector-like quarks) are given by operators of the form
\begin{align}\label{eq:GeneralCPViolation}
    \zeta_u \widetilde{H} \frac{\overline{q}_L\,\mathcal{C}\,\Sigma_u\,u_R}{f_u}\,,  \qquad \zeta_u' \widetilde{H} \frac{\overline{q}_L\, \mathcal{C}\,\textrm{det}(\Sigma_u)\, \widetilde{\Sigma}_u\, u_R}{f_u^5} \,, \;\;
\end{align}
with $\zeta_u$ and $\zeta_u'$ being some coefficients (encompassing, for example, loop factors) and we have the dimensionless quantity
\beqa
\mathcal{C} \equiv \prod_i \bigg(\frac{(\Sigma_u \Sigma_u^\dagger)^{u_i}}{f_u^{2 u_i}} \frac{(\widetilde{\Sigma}_u \widetilde{\Sigma}_u^\dagger)^{\tilde{u}_i}}{f_u^{4 \tilde{u}_i}} \frac{(\Sigma_d \Sigma_d^\dagger)^{d_i}}{f_u^{2 d_i}} \frac{(\widetilde{\Sigma}_d \widetilde{\Sigma}_d^\dagger)^{\tilde{d}_i}}{f_u^{4 d_i}} \bigg) ~.
\eeqa
Here, $u_i$, $\tilde{u}_i$, $d_i$, and $\tilde{d}_i$ are non-negative integers and $\widetilde{\Sigma}_{u,d}$ are defined as
\begin{align}\label{eq:TildeDefs}
    (\widetilde{\Sigma}_u)^\gamma_k \equiv \frac{1}{2} (\Sigma_u^*)_\alpha^i (\Sigma_u^*)_\beta^j \epsilon_{i j k}\epsilon^{\alpha \beta \gamma} = f_u^2\,V^\dagger_{\rm CKM} \,\cdot\,\textrm{diag} \bigg(\frac{m_c}{m_t}, \frac{m_u}{m_t}, \frac{m_c m_u}{m_t^2} \bigg),\\
    (\widetilde{\Sigma}_d)^C_k \equiv \frac{1}{2} (\Sigma_d^*)_A^i (\Sigma_d^*)_B^j \epsilon_{i j k}\epsilon^{A B C} = f_d^2 \, \textrm{diag} \bigg(\frac{m_s}{m_b}, \frac{m_d}{m_b}, \frac{m_s m_d}{m_b^2} \bigg), \nonumber
\end{align}
where $\epsilon$ denotes the 3-dimensional Levi-Civita symbol. Notably, all independent operators composed of the spurion VEV's $\Sigma_u$ and $\Sigma_d$ which might contain a complex phase can be constructed from these operators, up to real scalar factors such as $|\textrm{det}(\Sigma_u)|^2$ which don't contribute anything to the phase and have a trivial effect on the matrix structure.

We can actually simplify this structure further by noting that the matrices $\Sigma_{u,d}$ and those built out of their antisymmetric products, $\widetilde{\Sigma}_{u,d}$, are not independent. Specifically, we can note that
\begin{align}\label{eq:SigmaIdentity}
    \det (\Sigma_{u,d}) (\Sigma_{u,d})^{-1} = \widetilde{\Sigma}_{u,d}^\dagger ~.
\end{align}
So, we can rewrite the second class of operators given in Eq.~(\ref{eq:GeneralCPViolation}) as
\beqa
  &&  \zeta_u \widetilde{H}\overline{q}_L\, \frac{ \mathcal{C}\,\textrm{det}(\Sigma_u) \,\widetilde{\Sigma}_u}{f_u^5}  u_R = \zeta_u \widetilde{H} \overline{q}_L\mathcal{C} \bigg( \frac{\textrm{det}(\Sigma_u)\widetilde{\Sigma}_u (\Sigma_u)^{-1}}{f_u^4}\bigg) \frac{\Sigma_u}{f_u}  u_R = \zeta_u \widetilde{H} \overline{q}_L \mathcal{C} \bigg( \frac{\widetilde{\Sigma}_u \widetilde{\Sigma}_u^\dagger}{f_u^4}\bigg) \frac{\Sigma_u}{f_u} u_R \nonumber \\
&    \equiv& \zeta_u \widetilde{H}\, \overline{q}_L\, \mathcal{C}'\, \frac{\Sigma_u}{f_u}  u_R ~,
\eeqa
where in the final line we have simply reabsorbed the additional factor of $\widetilde{\Sigma}_u \widetilde{\Sigma}_u^\dagger/f_u^4$ into a product $\mathcal{C}'$, which, referring to Eq.~\eqref{eq:GeneralCPViolation}, can clearly be written as another possible CP-violating operator $\mathcal{C}$ in that equation. So, \emph{all} possible leading-order operators which might contribute to the strong CP phase in the up-like SM quark mass matrix are of the form
\begin{align}\label{eq:GeneralCPViolationFinal}
\zeta_u\,\widetilde{H} \frac{\overline{q}_L\,\mathcal{C}\,\Sigma_u\,u_R}{f_u}\,,
\end{align}
where $\mathcal{C}$ is the set of all operators that can be written in the form described in Eq.~\eqref{eq:GeneralCPViolation}.

Our task now becomes to identify \emph{which} operators of the form given in Eq.~\eqref{eq:GeneralCPViolationFinal} actually contribute to the strong CP phase. To do this, we consider a general operator of the form given in that equation, and determine its contribution to the strong CP phase. The determinant of the up-like SM quark mass matrix after including this operator is
\begin{align}\label{eq:uplikeCPPhase}
    \frac{y_u^3\, v^3}{2\sqrt{2} \, \Lambda^3} \det\left[(\mathbf{1} + \tilde{\zeta}_u\, \mathcal{C})\Sigma_u \right] = \frac{y_u^3 \,v^3}{2\sqrt{2}\, \Lambda^3} \det (\Sigma_u) \det(\mathbf{1} + \tilde{\zeta}_u\,\mathcal{C}) ~, \;\;\; \tilde{\zeta}_u \equiv \frac{\zeta_u f_u }{y_u \Lambda} ~,
\end{align}
where $\mathbf{1}$ is the $3 \times 3$ identity matrix, while we remind the reader that $\Lambda \sim f_u$ is the scale characteristic of the dimension-5 operators which grant the SM quarks mass.~\footnote{One should note that the second step in Eq.~(\ref{eq:uplikeCPPhase}) is predicated on the invertibility of $\Sigma_u$---if $\Sigma_u$ is singular (for example, if the up quark is massless), the determinant on the left side of Eq.~(\ref{eq:uplikeCPPhase}) vanishes and therefore the CKM matrix has a vanishing contribution to the strong CP phase, as we'd anticipate since the strong CP phase is unphysical in this case.} Since we are working in a basis in which $\det (\Sigma_u)$ and all of the constant prefactors of this expression are real, the strong CP phase contribution from the operator $\mathcal{C}$ is
\begin{align}
    \arg \left[\det ( \mathbf{1} + \tilde{\zeta}_u \,\mathcal{C} )\right] = \arg \left\{ 1 + \tilde{\zeta}_u\, \mbox{Tr} \, \mathcal{C} + \frac{1}{2}\, \tilde{\zeta}_u^2\, \left[ (\mbox{Tr}\,\mathcal{C})^2 - \mbox{Tr} \, \mathcal{C}^2 \right] + \frac{1}{6} \,\tilde{\zeta}_u^3\, \det(\mathcal{C})  \right\} ~.
\end{align}
The spurion product $\mathcal{C}$ will always be small, so we can expand to leading order in $\mathcal{C}$ and thus find that
\begin{align}\label{eq:DetToTrace}
    \arg \left[\det ( \mathbf{1} + \tilde{\zeta}_u \,\mathcal{C} )\right] \approx \arg (1 + \tilde{\zeta}_u \,\mbox{Tr} \, \mathcal{C} ) \approx \tilde{\zeta}_u\, \mbox{Im}( \mbox{Tr}\, \mathcal{C}) ~.
\end{align}
Note that if a given operator $\mathcal{C}$'s contribution to the strong CP phase vanishes at $O(\mathcal{C})$ but appears in $O(\mathcal{C}^2)$ terms, one could just treat $\mathcal{C}^2$ as the effective $\mathcal{C}$ operator.
So, our task has become identifying those spurion products $\mathcal{C}$ of the form specified in Eq.~(\ref{eq:GeneralCPViolation}) with complex traces, and estimating their magnitude.

First, it is useful to recast the spurion products $\mathcal{C}$ in a somewhat more helpful form. In the basis we have chosen, $\Sigma_d$ and $\widetilde{\Sigma}_d$ are real and diagonal, while $\Sigma_u$ and $\widetilde{\Sigma}_u$ are products of real diagonal matrices with the CKM matrix. We can therefore generically write the spurion products $\mathcal{C}$ described in Eq.~(\ref{eq:GeneralCPViolation}) as
\begin{align}
    \mathcal{C} = \prod_i ( U_i\,V_{\rm CKM} \,D_i \,V_{\rm CKM}^\dagger ) ~,
\end{align}
where each $D_i$ is a real diagonal matrix made up of products of $\Sigma_d \Sigma_d^\dagger$ and $\widetilde{\Sigma}_d \widetilde{\Sigma}_d^\dagger$ (with appropriate powers of $f_u$ included), while $U_i$ is a real diagonal matrix made up of products of diagonalized $\Sigma_u \Sigma_u^\dagger$ and $\widetilde{\Sigma}_u \widetilde{\Sigma}_u^\dagger$, that is, the up-like $\Sigma_u$ matrices if $V_{\rm CKM}$ is taken to be the identity matrix. To isolate the CP-violating part of a given operator, it's most convenient to work in a basis in which we consider the spurion quantity
\begin{align}
    \mathcal{C}^{\rm I} \equiv \mathcal{C}-\mathcal{C}^\dagger,
\end{align}
which are obviously purely CP-odd. It is straightforward to verify explicitly that for \emph{any} collection of real diagonal $U_i$'s and $D_i$'s, the CP-odd combination $\mathcal{C}^{\rm I}$ vanishes unless it features at least four CKM matrices (``coincidentally'', the number of different CKM matrix elements needed to form the Jarlskog invariant). So, we can begin our exploration of the CP-violating operators with those spurion products that feature exactly four CKM matrix insertions. The simplest CP-odd spurion products with four insertions can be written in the form
\begin{align}\label{eq:LOSpurions}
    \mathcal{C}_A = U_1 \,V_{\rm CKM} \,D_1 \,V_{\rm CKM}^\dagger\, U_2\, V_{\rm CKM}\, D_2\, V_{\rm CKM}^\dagger, \;\;\; \quad\mathcal{C}^{\rm I}_A \equiv \mathcal{C}_A - \mathcal{C}^\dagger_A,
\end{align}
for which we find that
\beqa\label{eq:CPViolationLO}
   && \mbox{Im}( \mbox{Tr} \, \mathcal{C}^{\rm I}_A) = 2 J \, \mathcal{F}(U_1,U_2)\, \mathcal{F}(D_1,D_2)~,\\
   &\mbox{with}& \mathcal{F}(A,B) \equiv [A_{33}B_{11}-A_{11}B_{33} + A_{22}B_{33}-A_{33}B_{22} + A_{11}B_{22}-A_{22}B_{11}]~,\nonumber
\eeqa
where $J$ is the Jarlskog invariant. Because we are only concerned with the trace of these spurion products, we see that $\mathcal{C}_A$ as described in Eq.~(\ref{eq:LOSpurions}) encapsulates all possible contributions to the strong CP phase for any spurion product featuring four CKM matrix insertions, due to the cyclic property of the trace. Notably, the expression in Eq.~(\ref{eq:CPViolationLO}) vanishes whenever either $U_1 = U_2$ or $D_1 = D_2$. We can straightforwardly classify all possible spurion insertions of the form given in Eq.~(\ref{eq:LOSpurions}). By inserting the definitions of the $\Sigma$ fields given in Eqs.~(\ref{eq:SigmaDefs}) and (\ref{eq:TildeDefs}), and leveraging the identity in Eq.~(\ref{eq:SigmaIdentity}), we see that if $U_1$ and $U_2$ are only constructed out of diagonalized products of $\Sigma_u \Sigma_u^\dagger$ and $\widetilde{\Sigma}_u \widetilde{\Sigma}_u^\dagger$, and $D_1$ and $D_2$ are only constructed out of the products of the diagonalized $\Sigma_d \Sigma_d^\dagger$ and $\widetilde{\Sigma}_d \widetilde{\Sigma}_d^\dagger$, then all $\mathcal{F}(U_1,U_2)$ and $\mathcal{F}(D_1,D_2)$ will take the form
\begin{align}
    &\mathcal{F}(U_1,U_2) = \bigg( \frac{m_u m_c}{m_t^2} \bigg)^{2 \gamma} F_{\alpha,\beta} \bigg( \frac{m_u}{m_t}, \frac{m_c}{m_t}\bigg),\;\;\; \mathcal{F}(D_1,D_2) = \bigg( \frac{m_d m_s f_d^3}{m_b^2 f_u^3} \bigg)^{2\sigma} \bigg(\frac{f_d}{f_u} \bigg)^{2(\delta+\rho)} F_{\delta,\rho} \bigg( \frac{m_d}{m_b}, \frac{m_s}{m_b}\bigg),\\
    &F_{\alpha,\beta}(A,B) \equiv \bigg[ A^{2 \alpha} - B^{2 \alpha} + B^{2 \beta} - A^{2 \beta} + A^{2 \beta} B^{2 \alpha} - B^{2 \beta} A^{2 \alpha} \bigg],\nonumber
\end{align}
where $\alpha,\beta,\gamma,\delta,\rho,\sigma$ are all non-negative integers. The maximum magnitude of $\mathcal{F}(U_1,U_2)$ occurs when $\alpha = 1$ and $\beta \geq 2$ (or vice versa) and $\gamma=0$, while the maximum magnitude of $\mathcal{F}(D_1,D_2)$ occcurs when $\delta=1$ and $\rho=2$ (or vice versa) and $\sigma=0$, up to small corrections proportional to squares of quark mass hierarchies, so we see that for any $\mathcal{C}_A$,
\begin{align}
    \tilde{\zeta}_u \, \left|\mbox{Im}(\mbox{Tr}\,\mathcal{C}^{\rm I}_A)\right| \lesssim 2 A^2\, \eta\, \lambda^6\, \tilde{\zeta}_u\, \frac{f_d^6}{f_u^6} \frac{m_s^2 m_c^2}{m_b^2 m_t^2} \sim (3.1 \times 10^{-13})\, \tilde{\zeta}_u\, \frac{f_d^6}{f_u^6}~,
\end{align}
where $A$, $\eta$, and $\lambda$ are the usual Wolfenstein parameters~\cite{Wolfenstein:1983yz}. Up to mildly differing conventions (assuming, for example, that the UV cutoff is $f_u$ instead of $\Lambda$), this agrees with our sample numerical result in Eq.~(\ref{eq:leading-CP-op-numerics}), as can be seen by leveraging the fact that $f_d/f_u = y_u m_b/(y_d m_t)$. In general, spurion products featuring more insertions of the CKM matrix should be even smaller, since they must feature additional insertions of $\Sigma_{u,d}$ and/or $\widetilde{\Sigma}_{u,d}$. As discussed in the main text, the ratio $f_d/f_u$ can feasibly vary between $f_d/f_u \sim m_b/m_t$ and $f_d/f_u \sim 1$, indicating that for $\tilde{\zeta}_u \sim 1$, our estimated higher-order corrections to the strong CP phase range between $O(10^{-13})$ and $O(10^{-24})$---even on the high end of this range, these corrections are well below the level constrained by experiment.

\section{Detailed Calculation for the Thermal History of GBs}
\label{sec:thermal-GB}
In this Appendix, we present our calculation of the effective radiative degrees of freedom at recombination in greater detail than in the main text. From Eq.~(\ref{eq:delta-n-eff-formula}), we know that in order to compute $\Delta N_{\rm eff}$, we only need to evaluate the relic abundances for all 26 Goldstone bosons in the model. From the main text, we recall that this is achieved by solving Eq.~(\ref{eq:boltzmann-eq}). Recasting this equation in terms of the number density scaled by the entropy density of the universe, we can put the Boltzmann equations of Eq.~(\ref{eq:boltzmann-eq}) into a somewhat more manageable form,
\begin{align}\label{eq:boltzmann-eq-Y}
    \frac{d Y_a}{d q} = -\ln(10) \frac{\sqrt{90}\,M_{\rm pl}}{\pi \sqrt{g_{*}(T)} T^2}\bigg( 1 + \frac{1}{3} \frac{d \ln(g_{*s}(T))}{d \ln(T)}\bigg) \Gamma_a (q) \bigg(\frac{45 \,\zeta(3)}{2 \pi^4 g_{*s}(T)}-Y_a(q) \bigg)~,
\end{align}
where $g_{*s}(T)$ and $g_{*}(T)$ are the effective relativistic degrees of freedom of the thermal bath for entropy and energy, respectively, $M_{\rm pl} = 2.435 \times 10^{18} \; \rm GeV$ is the reduced Planck mass, and $q \equiv \log_{10}(T/\mbox{GeV})$.
We can then evaluate the relic yield of each Goldstone boson by numerically solving Eq.~(\ref{eq:boltzmann-eq-Y}), starting at the reheating temperature $T_{\rm reh}$ and evolving down to some low temperature below which all Goldstone bosons are reliably decoupled from the SM bath (we find that $T = 10^7 \; \textrm{GeV}$, or equivalently $q = 7$, is sufficient for our purposes). We note that in the form of Eqs.~(\ref{eq:boltzmann-eq}) and (\ref{eq:boltzmann-eq-Y}), we have already made a significant assumption: specifically, we have assumed that the relic abundance of the Goldstone bosons will be dominated by processes involving only a single Goldstone boson coupled to SM particles or the vector-like quarks, that is, those particles which are in equilibrium with the SM thermal bath. As long as the SM bath temperature $T$ is far lower than the decay constant $f_a$ of a Goldstone boson $a$, as is usually assumed in, \eg, \cite{Baumann:2016wac,Green:2021hjh,Masso:2002np}, this is a perfectly reasonable assumption, since the thermal rates of processes with a larger number of Goldstone bosons will generically suffer at least $T^2/f_a^2$ suppression relative to the thermal rates of single-Goldstone processes. In our scenario, however, the hierarchies of different Goldstone bosons' decay constants will lead to the somewhat troubling circumstance of evaluating Goldstone production rates at temperatures far in excess of their decay constants, where perturbativity will break down altogether. As it turns out, this concern is unfounded: in all parameter space points we consider processes featuring a single Goldstone boson are rapid enough to keep a given GB in thermal equilibrium with the SM until temperatures $T \lesssim f_a$, except in cases in which the decay constant $f_a$ is larger than the highest reheating temperature we consider, $T=10^{15}\; \textrm{GeV}$, which as we note in the main text is close to the largest reheating temperature that is realistically achievable. More troublingly, however, we will still need to evaluate rates at temperatures comparable to Goldstone bosons' decay constants, at which perturbativity is maintained but the rates of processes with two Goldstone bosons, which may or may not be the same species, might be comparable to other $2 \rightarrow 2$ processes in which only a single GB is scattered. Numerically, however, we find that including these processes, while they can have an $O(1)$ effect on individual Goldstone boson production rates, yields only percent-level discrepancies in our final results.

Since, as we have argued, solving Eq.~(\ref{eq:boltzmann-eq-Y}) will allow us to accurately compute $\Delta N_{\rm eff}$, we must now discuss the evaluation of the various quantities appearing in the equation. We shall discuss these each in turn, beginning with the effective relativistic degrees of freedom $g_{*s}$ and $g_{*}$ and then discussing the production rate $\Gamma_a (T)$ for all processes relevant in the model.

\subsection{Effective Relativistic Degrees of Freedom $g_{*s}$ and $g_{*}$}

The effective relativistic degrees of freedom $g_{*s}(T)$ and $g_{*}(T)$ will include the contributions of both of the SM particles and the new fields we have introduced, namely the vector-like heavy quarks and the Goldstone bosons themselves. The SM thermal bath's contribution is easy to compute: because we find that for any point in parameter space which satisfies the nucleon coupling constraints fron SN 1987A, all Goldstone bosons decouple well before the electroweak and QCD phase transitions, both $g_{*s}$ and $g_{*}$ coming from the SM bath will be the usual 106.75 for temperature above the top quark mass. In the case of the vector-like heavy quarks, their electric charges ensure that they will remain in thermal equilibrium with the SM bath throughout the universe's history, so we can include the contribution of a vector-like quark of mass $m_{Q}$ to $g_{*s}(T)$ and $g_{*}(T)$ as~\cite{Husdal:2016haj}
\begin{align}
    \begin{matrix}
        g_{*s}^Q (T) = 12 \times \frac{15}{\pi^4} \int_{x_Q}^{\infty} \frac{(4 u^2-x_Q^2)\sqrt{u^2-x_Q^2}}{e^u+1} du, & g_{*}^Q (T) = 12 \times \frac{15}{\pi^4} \int_{x_Q}^{\infty} \frac{u^2\sqrt{u^2-x_Q^2}}{e^u+1} du, & x_Q \equiv \frac{m_Q}{T},
    \end{matrix}
\end{align}
where we have kept the factor of $12=3 \times 4$  for the degrees of freedom of one vector-like heavy quark explicitly.

For maximum precision, we would also have to include the contributions of the Goldstone bosons themselves to $g_{*s}$ and $g_{*}$. Naively we might expect that for scenarios in which a large number of Goldstone bosons achieve thermal equilibrium with the bath, the GB contributions to $g_{*s}$ and $g_{*}$, which might modify their values by as much as $\sim 30\%$, would similarly modify our results for $\Delta N_{\rm eff}$. However, we find that including the extra scalar radiative degrees of freedom result in only a percent level modification to our results for $\Delta N_{\rm eff}$, due to a remarkable cancellation in Eq.~(\ref{eq:delta-n-eff-formula}). We can explore this cancellation semi-analytically. First, assuming $T_a$ is the temperature of a given Goldstone boson $a$ (which may or may not be decoupled from the SM thermal bath), and $g_{*s}^0(T)$ and $g_{*}^0(T)$ are the effective relativistic entropy and energy degrees of freedom as a function of the photon temperature $T$, \emph{omitting} the contribution of the GB's, then for any photon temperature $T$ the \emph{total} effective relativistic degrees of freedom will be given by
\beqa\label{eq:g-exact-def}
    \begin{matrix}
        g_{*s}(T) = g_{*s}^0(T) + \sum_a \bigg( \dfrac{T_a}{T}\bigg)^3, & g_{*}(T) = g_{*}^0(T) + \sum_a \bigg( \dfrac{T_a}{T}\bigg)^4.
    \end{matrix}
\eeqa
The temperature $T_a$ of each Goldstone boson can be found by recalling that $Y_{a} = n_a/s$, where $n_a$ is the number density of the species $a$ and $s$ is the entropy density of the universe. In terms of the effective entropy radiative degrees of freedom of the universe, $g_{*s}(T)$, we have
\beqa \label{eq:eff-temp}
    \bigg( \frac{T_a}{T} \bigg)^3 = \frac{2 \pi^4}{45 \zeta(3)} g_{*s}(T) Y_a (T),
\eeqa
where $Y_a(T)$ is the yield $Y_a$ when the photon bath is at a temperature $T$ (which will be determined as the solution to the Boltzmann equations). Inserting Eq.~(\ref{eq:eff-temp}) into Eq.~(\ref{eq:g-exact-def}), we have
\beqa\label{eq:g-exact}
    \begin{matrix}
        g_{*s}(T) = \bigg(1- \dfrac{2 \pi^4}{45\, \zeta(3)} \sum_a Y_a (T) \bigg)^{-1} g_{*s}^0(T), & g_{*}(T) = g_{*}^0(T) + \sum_a \bigg( \dfrac{2 \pi^4}{45\, \zeta(3)} g_{*s}(T) Y_a(T) \bigg)^{4/3}.
    \end{matrix}
\eeqa
The dominant numerical effect on the results of our Boltzmann equations from including the Goldstone bosons in the calculation of $g_{*s}$ and $g_{*}$ is the modification of the yield $Y_a$ for a species $a$ in thermal equilibrium with the SM: we see from Eq.~(\ref{eq:g-exact}) and the Boltzmann equation that in the approximation that a given Goldstone boson $a$ decouples instantaneously at some temperature $T^{a}_{\rm dec}$, its relic yield $Y_{a, \infty}$ takes on a constant value given by
\begin{align}\label{eq:corrected-Y-inf}
    Y_{a, \infty} = \bigg( 1- \frac{2 \pi^4}{45\, \zeta(3)} \sum_b Y_b(T^a_{\rm dec}) \bigg)\,, \;\;\; Y^0_{a, \infty}  \equiv \frac{45\,\zeta(3)}{2 \pi^4 g_{*s}^0(T^a_{\rm dec})}~,
\end{align}
where $Y^0_{a, \infty}$ is the relic yield of the Goldstone boson in the same instantaneous decoupling approximation when we do \emph{not} include the GB's in our calculation of $g_{*s}$ and $g_{*}$, and the quantities $Y_b (T^a_{\rm dec})$ are, as notation suggests, the yields of the Goldstone species $b$ when the photon bath reaches the temperature $T^a_{\rm dec}$ at which $a$ decouples. So, we can see that generally, including the Goldstone bosons in our calculation of $g_{*s}$ and $g_{*}$ will increase the effective relativistic degrees of freedom at each Goldstone boson decoupling and reduce their relic yields---given the large number of Goldstone bosons we consider here (26 of them), we might find that a calculation including the Goldstone bosons in $g_{*s}$ and $g_{*}$ might reduce our predicted relic yields by as much as $\sim 20\%$ compared to a computation where they are omitted. However, an interesting cancellation occurs when we use these relic yields to compute $\Delta N_{\rm eff}$. Specifically, while the relic yields of the Goldstone bosons are modified from $Y^0_{a, \infty}$ according to the factor given in Eq.~(\ref{eq:corrected-Y-inf}), the effective relativistic degrees of freedom of the SM bath before neutrino decoupling, $g_{*s,\infty}$ is modified by a factor given in Eq.~(\ref{eq:g-exact}). So, inserting Eq.~(\ref{eq:corrected-Y-inf}) into Eq.~(\ref{eq:delta-n-eff-formula})  we see this quantity is given by (again assuming that all Goldstone bosons decouple instantaneously)
\begin{align}\label{eq:delta-n-modified}
    \Delta N_{\rm eff} = \sum_a \bigg[ \Delta N^0_{{\rm eff},a} \bigg( \frac{1-\frac{2 \pi^4}{45 \zeta(3)} \sum_b Y_b (T^a_{\rm dec})}{1-\frac{2 \pi^4}{45 \zeta(3)} \sum_b Y_{b,\infty}} \bigg)^{\frac{4}{3}} \bigg], \;\;\; \Delta N^0_{{\rm eff},a} \equiv \frac{4}{7} \bigg(\frac{11 }{4} \frac{ g^0_{*s,\infty}}{g^0_{*s}(T^a_{\rm dec})}\bigg)^{\frac{4}{3}},
\end{align}
where $\Delta N^0_{{\rm eff},a}$ is simply the contribution of a Goldstone boson species $a$ to $\Delta N_{\rm eff}$ if we had omitted the Goldstone bosons from our computation of $g_{*s}$ and $g_{*}$. In Eq.~(\ref{eq:delta-n-modified}), we see that the correction to $Y_{a, \infty}$ from including GB's in $g_{*s}$ and $g_{*}$ is partially cancelled by the correction to $g_{*s,\infty}$ from the same effect. In fact, because after decoupling the relic yield of a Goldstone boson remains constant, the correction factor to $\Delta N^0_{{\rm eff},a}$ in Eq.~(\ref{eq:delta-n-modified}) can only be unequal to unity in the event that some Goldstone boson species $b$ decouple at a lower temperature than $T^a_{\rm dec}$---otherwise the numerator and denominator of the correction cancel perfectly. The fact that the Goldstone bosons generically do decouple at different temperatures means that the cancellation is less than perfect, but numerically we find that the correction to the total value of $\Delta N_{\rm eff}$ is never greater than the percent level for any points in parameter space we consider. As such, we can ignore the effects of the Goldstone bosons in $g_{*s}$ and $g_{*}$ and compute these quantities including only the SM fields and the new vector-like heavy quarks, without fear of significant numerical inaccuracy.

\subsection{Computation of GB Production Rates}

For a given process with squared amplitude $|\mathcal{M}|^2$ that produces a single Goldstone boson of a species $a$, the production rate shall be~\cite{Green:2021hjh}
\beqa \label{eq:rate-master-equation}
    \Gamma_a = \frac{1}{n_a^{\rm eq}(T)} \int \prod_i \bigg( \frac{d^3 p_i \, f_i(p_i) }{(2 \pi)^3 2 E_i} \bigg) \prod_j \bigg( \frac{d^3 p_j \, [1 \pm f_j (p_j)]}{(2 \pi)^3 2 E_j} \bigg) |\mathcal{M}|^2 ~,
\eeqa
where $n_a^{\rm eq}(T) = \zeta(3) T^3/\pi^2$ is the equilibrium number density for a highly relativistic real scalar, $p_i$ denotes the momentum of the initial-state particles in the process, $p_j$ denotes the momentum of the final-state particles, $f_{i,j}$ denote the distribution functions for these particles in thermal equilibrium, with $[1 +(-) f_j]$ factors included for final-state particles for bosons (fermions), and $|\mathcal{M}|^2$ is evaluated by summing over all initial and final degrees of freedom. We note that the expression in Eq.~(\ref{eq:rate-master-equation}) assumes that the number density of the Goldstone boson $a$ is near its thermal equilibrium value---if instead the Goldstone boson number density is far lower than equilibrium, Eq.~(\ref{eq:rate-master-equation}) will overestimate the production rate by an $O(1)$ factor, due to Bose enhancement of the final state. In practice this has a negligible effect on our computation of $\Delta N_{\rm eff}$, since it will remain accurate for any Goldstone boson which ever equilibrates with the SM bath, and those GB's that \emph{never} achieve thermal equilibrium with the SM bath will have a negligible numerical effect on $\Delta N_{\rm eff}$, due to their small relic yields.

We can now use Eq.~(\ref{eq:rate-master-equation}) to find the rates of the various processes discussed in the main text and depicted in Figs.~(\ref{fig1}) and (\ref{fig2}), as well as those stemming from decay of vector-like heavy quarks. As noted in the main text, we can separate these processes into those which stem from the Goldstone boson couplings to SM quarks and those which stem from couplings to the heavy vector-like quarks. 

\subsubsection{SM Quark Interactions}

We shall begin with the processes from SM quark interactions with the Goldstone bosons. For a Goldstone boson $a$, these interactions are given by
\beqa\label{eq:generic-SM-couplings}
    \mathcal{L} \supset - \frac{\partial_\mu a}{f_u} \overline{u}_i \gamma^\mu \left[P_L (g_L^u)_{ij}+P_R (g_R^u)_{ij}\right] u_j - \frac{\partial_\mu a}{f_u} \overline{d}_i \gamma^\mu \left[P_L (g_L^d)_{ij}+P_R (g_R^d)_{ij}\right] d_j ~,
\eeqa
where $g^{u,d}_{L,R}$ are coupling matrices which can be derived following the procedure outlined in Section \ref{sec:GB}, and $i$ and $j$ are flavor indices. Instead of directly evaluating the integral in Eq.~(\ref{eq:rate-master-equation}), we can greatly simplify our work by making a handful of observations. First, we find that for points in parameter space which satisfy the SN 1987A constraints [\ie, Eq.~(\ref{eq:fu-fd-SN-constraints})], Goldstone bosons in the model will decouple long before the electroweak phase transition (EWPT). As a result, we can easily work in the massless limit for the external quarks and follow the analysis of \cite{Baumann:2016wac}, where the authors have replaced the full expression of Eq.~(\ref{eq:rate-master-equation}) with a simplified expression based on the center-of-mass frame cross section, which proves to be highly accurate in the limit that all incoming and outgoing particles are ultra-relativistic. From that work, we have
\beqa\label{eq:rate-SM-quark-equation}
    \Gamma_a^{\rm SM}(T) \approx \frac{1}{n_a^{\rm eq}(T)} \int \frac{d^3 p_1}{(2 \pi)^3} \frac{d^3 p_2}{(2 \pi)^3} \frac{f_1(p_1)}{2 E_1} \frac{f_2(p_2)}{2 E_2} [1 \pm f_3][1 \pm f_4]\,2\,s\, \sigma_{\textrm{cm}}(s) ~,
\eeqa
where $s$ is the usual Mandelstam variable, $\sigma_{\textrm{cm}}$ is the center-of-mass frame cross section $f_{1,2}$ are the distribution functions of the initial states, and $f_{3,4}$ are the distribution functions for the final states. Note that just as we do not average over the initial degrees of freedom in computing the squared matrix elements of Eq.~(\ref{eq:rate-master-equation}), we do not do so when computing $\sigma_{\textrm{cm}}$ here. The factor $[1 \pm f_3][1 \pm f_4]$ is given by
\beqa
    [1 \pm f_3][1 \pm f_4] \equiv \frac{1}{2}[1 \pm f_3(p_1)][1 \pm f_4(p_2)] + (p_1 \leftrightarrow p_2) ~,
\eeqa
with a $+(-)$ factor for a boson(fermion) state---apart from being more accurate than omitting Bose enhancement and Pauli blocking from the equation altogether, this factor also ensures that the same rate is computed regardless of whether we compute the forward or backward rate. In the massless limit, we can easily sum the cross sections of the processes of Figure \ref{fig1} over the flavor indices $i$ and $j$. Summing over both fermion and antifermion scattering as well as all quark flavors (including, for example, both up-like and down-like external quarks in the processes of Figures \ref{fig1}(c) and \ref{fig1}(d)), we find for a Goldstone boson species $a$ with couplings to SM quarks given as in Eq.~(\ref{eq:generic-SM-couplings}), the center-of-mass cross sections for the processes depicted in Figure \ref{fig1} are all identical, given by
\beqa\label{eq:light-quark-cross-sections}
    \sigma^{(a-d)}_{\textrm{cm}} &=& \frac{3}{8 \pi f_u^2} \frac{m_t^2}{v^2}\bigg(G^{u}_{\rm SM} + \frac{m_b^2}{m_t^2} G^{d}_{\rm SM}\bigg) ~, \\ 
    &&G^{u,d}_{\rm SM} \equiv \mbox{Tr} [(g^{u,d}_L\, R_{u,d} - R_{u,d} \,g^{u,d}_R)(R_{u,d} \,g^{u,d}_L - g^{u,d}_R\, R_{u,d}) ~. \nonumber 
\eeqa
It is important to note that, because the coupling matrices $g^{u,d}_L$ are Hermitian, these cross sections will always be real. Since the results of Eq.~(\ref{eq:light-quark-cross-sections}) have no dependence on the center-of-mass energy, and the processes depicted in Figures \ref{fig1}(a) and \ref{fig1}(c) feature two initial-state fermions and two final-state bosons while the processes in \ref{fig1}(b) and \ref{fig1}(d) feature one fermion and one boson in both the initial and final states, we can rewrite Eq.~(\ref{eq:rate-SM-quark-equation}) as
\beqa
    \Gamma^{\rm SM}_a (T) \approx \frac{49 \zeta(3) T^3}{64 \pi^2} \Bigl(\sigma^{(a)}_{\rm cm} + \sigma^{(c)}_{\rm cm}\Bigr) + \bigg( \frac{\pi^2 T^3}{144 \zeta (3)} + \frac{49 \zeta (3) T^3}{128 \pi^2} \bigg) (\sigma^{(b)}_{\rm cm} + \sigma^{(d)}_{\rm cm}) ~,
\eeqa
leading to the result of Eq.~(\ref{eq:SM-quarks-rate}) for this rate, given in the main text.

Before moving on, we note that it is a priori possible that after the EWPT, SM quark processes such as $c + g \rightarrow c + a$, which will produce a Goldstone boson $a$ at a rate which scales more slowly with temperature than the Hubble rate, will cause $a$ to ``freeze in'' at some lower temperature, a possibility discussed in, \eg, \cite{Baumann:2016wac,Green:2021hjh}. However, it was found in \cite{Green:2021hjh} that for heavier SM quarks, the decay constants for the Goldstones must be many orders of magnitude smaller than anything which will satisfy our SN 1987A constraint in order to effect any observable change in $\Delta N_{\rm eff}$ from this freeze-in, and while the calculation has not to our knowledge been carried out for quarks with masses below the temperature of the QCD phase transition, following the pattern of the heavier quark flavors it is unlikely that this freeze-in is feasible in our model for $f_u$ and $f_d$ values which satisfy the harsh constraints from SN 1987A. So, the rate computed in Eq.~(\ref{eq:SM-quarks-rate}) represents to an excellent approximation all Goldstone bosons' production rates stemming from processes featuring SM quarks.

\subsubsection{Vector-like Quark Interactions}

In direct analogy to the SM quarks, we write the coupling terms of some Goldstone boson $a$ to the vector-like heavy quarks $B$ and $T$ as
\beqa
\label{eq:generic-VL-couplings}
    \mathcal{L} \supset - \frac{\partial_\mu a}{f_u} \overline{T}_i \gamma^\mu \left[P_L (g_L^T)_{ij}+P_R (g_R^T)_{ij}\right] T_j - \frac{\partial_\mu a}{f_u} \overline{B}_i \gamma^\mu \left[P_L (g_L^B)_{ij}+P_R (g_R^B)_{ij}\right] B_j ~.
\eeqa
We note that by construction, $g^{T,B}_{L, R} = -(g^{u,d}_{L,R})^*$, however for clarity we denote them separately here. In stark contrast to Goldstone boson production from SM quark interactions, production from couplings to heavy vector-like quarks will generally occur at temperatures comparable to the heavy quark masses. As a result, the ultra-relativistic approximation we employed when computing the rate from SM quark interactions will not be applicable here. Instead, we must resort to using the master formula Eq.~(\ref{eq:rate-master-equation}). For the decay of a vector-like quark $Q_i \rightarrow Q_j + a$, depicted in Figure \ref{fig2}(a), we find using the full Fermi-Dirac and Bose-Einstein distributions for all involved fields that the rate is given by
\begin{align}\label{eq:1to2-rate-VLQs}
    &\Gamma_{Q_i \rightarrow Q_j + a}(T) =3 \frac{m_{Q_i}^3}{f_u^2} \bigl(|(g^Q_L)_{ij}|^2 + |(g^Q_R)_{ij}|^2\bigr)\gamma_{Q_i \rightarrow Q_j + a}  \bigg( \frac{m_{Q_j}}{T} \bigg)~,\\
    &\gamma_{Q_i \rightarrow Q_j + a}(x) \equiv \int_1^\infty d \epsilon \, \frac{e^{\epsilon x} x^2 [\log (1-e^{\epsilon +\sqrt{\epsilon^2-1}})-\log (1-e^{\epsilon -\sqrt{\epsilon^2-1}})-x \sqrt{\epsilon^2-1}]}{32 \pi \zeta(3) (1 + e^{\epsilon x})^2}\nonumber ~,
\end{align}
where $Q=T,B$ and the limit is taken such that $m_{Q_j} \ll m_{Q_i}$, which, thanks to the hierarchy of SM quark masses, is always a valid approximation to make here. We have kept the color factor, 3, explicit in the above expression. The analogous process involving antiparticles, $\overline{Q}_i \rightarrow \overline{Q}_j + a$, will have the same rate. The function $\gamma(m_{Q_i}/T)$ is exponentially suppressed for $T \ll  m_{Q_i}$ and has a power-like behavior $\gamma(m_{Q_i}/T) \propto (m_{Q_i}/T)^\eta$ with the power $\eta$ range from 0.75 to 0.94 for $m_{Q_i}/T$ from 0.1 to $10^{-6}$. It has a maximum value of 0.85 at $m_{Q_i}/T = 0.97$. In contrast to the production rates from SM quark interactions, then, for certain temperature regions the production rate of Goldstone bosons from this process will scale more slowly with temperature than the Hubble rate, even decreasing with increasing temperature when $T \gtrsim m_{Q_i}$. As a result, in contrast to the SM quark-mediated production, for which a Goldstone boson must be thermally coupled immediately after reheating if it is to be thermally coupled at all, the process of vector-like heavy quark decay can allow for Goldstone bosons to ``freeze in'' at later times. We shall see this phenomenon again in the case of $2 \rightarrow 2$ vector-like quark scattering as well---in fact, the processes which allow for the possibility of Goldstone bosons freezing in are perfectly analogous to those which appear in \cite{Baumann:2016wac,Green:2021hjh} at a much lower scale from interactions of Goldstone bosons with the SM quarks.

Performing the rate computation from Eq.~(\ref{eq:rate-master-equation}) for the $2 \rightarrow 2$ quark scattering processes depicted in Figures \ref{fig2}(b) and \ref{fig2}(c) is considerably more involved. However, we can attain an acceptable degree of accuracy with much simpler integrals if we evaluate these processes' rates using classical Boltzmann statistics, rather full quantum statistics. Comparing the results for the Boltzmann distributions with a more exact treatment accounting for the full quantum statistics, we find that using the Boltzmann distribution tends to underestimate the production rates by no more than $\sim 30\%$ (and frequently much less), which we deem sufficiently accurate for our purposes.~\footnote{A reader may be concerned that our use of quantum statistics in calculating the production rate from vector-like quark decay and Boltzmann statistics in calculating the rates of other processes featuring the vector-like quarks. Our motivations here are purely numeric: in the case of decay, the use of Boltzmann statistics differs from the quantum result by as much as an order of magnitude at temperatures significantly greater than the decaying vector-like quark's mass, so full quantum statistics is necessary to hold our estimated rates to a reasonable degree of accuracy. Since no such discrepancy appears for any of the processes in Figure \ref{fig2}, we have simplified our calculation considerably by evaluating the rates using Boltzmann statistics.} For the processes in Figures \ref{fig2}(b) and \ref{fig2}(c), we have
\beqa \label{eq:2to2-rates-VLQs}
\Gamma^a_{Q_i \overline{Q}_i \rightarrow a g}(T) = |(g^Q_L-g^Q_R)_{ii}|^2 \frac{m_{Q_i}^3 \alpha_s (T) }{f_u^2} \gamma_{Q_i \overline{Q}_i \rightarrow a g} \bigg( \frac{m_{Q_i}}{T}\bigg)~, \nonumber\\
\Gamma^a_{Q_i g \rightarrow Q_i a}(T) = |(g^Q_L-g^Q_R)_{ii}|^2 \frac{m_{Q_i}^3 \alpha_s (T)}{f_u^2} \gamma_{Q_i g \rightarrow a Q_i} \bigg( \frac{m_{Q_i}}{T}\bigg)~,
\eeqa
where $\alpha_s (T)$ is the strong force fine structure constant run in the $\overline{\rm MS}$ scheme up to the temperature $T$ (with the computation done using RunDec \cite{Chetyrkin:2000yt}) and the $\gamma$ functions are given by
\beqa \label{eq:2to2-gamma-functions}
&\gamma_{Q_i \overline{Q}_i \rightarrow a g}(x) = \dfrac{4 x^2}{\pi^2 \zeta(3)} \int_1^\infty d \epsilon \, \sqrt{\epsilon}K_{1}(2 x \sqrt{\epsilon}) \arctanh \bigg( \sqrt{1-\frac{1}{\epsilon}} \bigg)~, \nonumber\\
&\gamma_{Q_i g \rightarrow a Q_i}(x) = \dfrac{x^2}{16 \pi^2 \zeta(3)} \int_1^\infty d\epsilon \, \epsilon^{-\frac{3}{2}} \bigg( 1 - \frac{1}{\epsilon} \bigg) K_1 (x \sqrt{\epsilon}) [1 - 4 \epsilon + 3 \epsilon^2 - 2 \epsilon^2 \log(\epsilon)]~.
\eeqa
We note that the rate $\Gamma^a_{Q_i g \rightarrow Q_i a}(T)$ is also accompanied by an identical rate from antiquark scattering. In principle, there are also analogous processes to those  featuring the SM hypercharge gauge boson in lieu of the gluon---these may be extracted from the rates above by making the substitution $\alpha_s \rightarrow \alpha_Y/4$ times the square of hypercharge, where $\alpha_Y$ is the SM hypercharge fine structure constant. However, we find that because of the color factor and the discrepancy between $\alpha_s$ and $\alpha_Y$ (even accounting for renormalization group equation up to the high temperatures at which scattering takes place), the Goldstone boson production rate from the SM hypercharge gauge boson scattering does not exceed $\sim 5 \%$ of the rate from gluon scattering.

\section{Analogous Setup with Renormalizable Couplings}\label{sec:ren-model}

In our model thus far, we have employed a dimension-5 operator in order to realize the appropriate cancellation between the strong CP phase coming from the SM quarks' mass matrices and that which emerges from our vector-like heavy quarks $B$ and $T$. We have made this choice for the sake of simplicity, but a reader may be concerned that such a construction might be difficult to achieve with solely renormalizable coupling terms, or what phenomenological differences a renormalizable construction might exhibit compared with the setup we've previously presented. To address both of these questions, in this appendix we present a modified version of the model which only relies on renormalizable couplings at tree level, demonstrate that it exhibits a similar suppression of higher-order corrections to the strong CP phase as our non-renormalizable model, and briefly comment on how the phenomenology of the renormalizable construction might compare to that of the non-renormalizable one.

\begin{table}[hb!]
	\centering
	\renewcommand{\arraystretch}{1.4}
		\begin{tabular}{ c | c | c | c | c |c |c }
			\hline \hline
  &   $[SU(3)_c \times SU(2)_W \times U(1)_Y] $ & $SU(3)_{q_L}$ & $SU(3)_{d_R}$ & $SU(3)_{u_R}$ & $U(1)_u$ & $U(1)_d$ \\ \hline
  $U_L$ & $(3, 1)_{2/3}$ & 3 & 1 & 1 & 0 &  0 \\ \hline 
  $U_R$ & $(3, 1)_{2/3}$ & 3 & 1 & 1 & 0 & 0 \\ \hline 
  $D_L$ & $(3, 1)_{-1/3}$ & 3 & 1 & 1 & 0 & 0 \\ \hline
  $D_R$ & $(3, 1)_{-1/3}$ & 3 & 1 & 1 & 0 & 0 \\ \hline
  \hline
		\end{tabular}
	\caption{
	Additional matter content of a renormalizable version of the construction presented in the main text, added onto the quark and scalar fields listed in Table \ref{tab:model}. Note that the fields $U$ and $D$ are simply vector-like quarks with the same representations under the MFV group as the left-handed quarks in Table \ref{tab:model}.} \label{tab:ren-model}
\end{table}

In our renormalizable setup, the particle content listed in Table \ref{tab:model} is extended to include additional flavor triplets of up-like and down-like vector-like quarks, $U_{L,R}$ and $D_{L,R}$, with representations under the SM and MFV groups given in Table \ref{tab:ren-model}. Incorporating the particle content of this Table with the existing fermion and scalar content given in Table \ref{tab:model}, we find that at tree level, the mass terms of Eq.~(\ref{eq:fermion-mass-terms}) are replaced with
\begin{align}
    \mathcal{L}_F \supset &-y_d \overline{q}_L H d_R - y_D \overline{D}_L \Sigma_d d_R - M_d \overline{D}_L D_R - \eta_d \overline{B}_L \Sigma_d^* B_R \\
    &-y_u \overline{q}_L \widetilde{H} U_R - y_U \overline{U}_L \Sigma_u u_R - M_u \overline{U}_L U_R -\eta_u \overline{T}_L \Sigma_u^* T_R\nonumber + h.c.,
\end{align}
where $y_{d,u}$, $y_{D,U}$, $M_{d,u}$, and $\eta_{d,u}$ are all real constants. Separating our fermions into groups of down-like quarks $(d,D,T)$ and up-like quarks $(u,U,T)$, we find that after electroweak symmetry breaking, the quarks have mass matrices given by
\begin{align}
    \begin{matrix}
    \mathcal{M}^d = \begin{pmatrix}
    0 & \frac{y_d v}{\sqrt{2}} & 0\\
    y_D \langle \Sigma_d \rangle & M_d & 0\\
    0 & 0 & \eta_d \langle \Sigma_d \rangle^*
    \end{pmatrix}, & \mathcal{M}^u = \begin{pmatrix}
    0 & \frac{y_u v}{\sqrt{2}} & 0\\
    y_U \langle \Sigma_u \rangle & M_u & 0\\
    0 & 0 & \eta_u \langle \Sigma_u \rangle^*
    \end{pmatrix}.
    \end{matrix}
\end{align}
It is straightforward to use block matrix determinant identities to derive that
\begin{align}\label{eq:ren-mass-matrices}
    \begin{matrix}
    \det(\mathcal{M}^d) = -\dfrac{y_d^3 y_D^3 \eta_d^3 v^3}{2 \sqrt{2}} |\det(\langle \Sigma_d \rangle)|^2, & \det(\mathcal{M}^u) = -\dfrac{y_u^3 y_U^3 \eta_u^3 v^3}{2 \sqrt{2}} |\det(\langle \Sigma_u \rangle)|^2 ~.
    \end{matrix}
\end{align}
So, as long as the product $y_d y_D \eta_d y_u y_U \eta_u >0$, the strong CP problem is solved at tree level in this arrangement. To leading order in $v$, the mass matrices in Eq.~(\ref{eq:ren-mass-matrices}) can be diagonalized so that
\begin{align}
\begin{matrix}
    \mathcal{M}^d_{\textrm{diag}} = \begin{pmatrix}
    \frac{y_d v}{\sqrt{2}}\frac{y_D f_d}{\sqrt{M_d^2 + y_D^2 f_d^2 R_d^2}} R_d & 0 & 0\\
    0 & \sqrt{M_d^2+y_D^2 f_d^2 R_d^2} & 0\\
    0 & 0 & \eta_d f_d R_d
    \end{pmatrix},\vspace{0.4cm}\\ 
    \mathcal{M}^u_{\textrm{diag}} = \begin{pmatrix}
    \frac{y_u v}{\sqrt{2}}\frac{y_U f_u}{\sqrt{M_u^2 + y_U^2 f_u^2 R_u^2}} R_u & 0 & 0\\
    0 & \sqrt{M_u^2+y_U^2 f_u^2 R_u^2} & 0\\
    0 & 0 & \eta_u f_u R_u
    \end{pmatrix},
\end{matrix}
\end{align}
where $f_{u,d}$ and $R_{u,d}$ are defined as in the main text.~\footnote{Note that because one of the matrices used to diagonalize these mass matrices has a determinant of -1, the transformation to mass eigenstates does not preserve the determinant of the original mass matrix.} We can then see that for $y_{U,D} \sim 1$, then $M_{u,d} \sim f_{u,d}$ will preserve the hierarchical structure of the SM quark masses, up to $O(1)$ adjustments to the diagonal values of the $R_u$ and $R_d$ matrices from the setup in the non-renormalizable theory. We also note that in this case, all of the newly-introduced vector-like $U$ quarks will acquire masses comparable to the scale $f_u$, while all of the $D$ quarks will acquire masses comparable to $f_d$, in stark contrast to the hierarchical structure observed in the $B$ and $T$ quarks.

Using similar operator analysis techniques as we have previously used in the main text and Appendix \ref{sec:more-operators}, we can further establish that this renormalizable setup lacks significant contributions to the strong CP phase from higher-order operators. First, we note that the $(1,3)$, $(3,1)$, and $(2,3)$ blocks of the mass matrices in Eq.~(\ref{eq:ren-mass-matrices}) have vanishing radiative corrections for precisely the same reason that the off-diagonal blocks of the non-renormalizable mass matrix vanish: Any operator which might contribute to these blocks must have a bi-triplet or bi-anti-triplet structure in $SU(3)_Q \times SU(3)_d \times SU(3)_u$, which is impossible to construct using the triplet-anti-triplet scalar VEV's $\Sigma_u$ and $\Sigma_d$. A similar argument also holds to enforce the vanishing of the $(3,2)$ block of the mass matrix. A contribution to this block must have the $SU(3)_Q \times SU(3)_d \times SU(3)_u$ representation of $(3,1,1)$ or $(\overline{6},1,1)$. If we have an operator containing $x$ $\Sigma_d$, $y$ $\Sigma_d^*$, $z$ $\sigma_u$ , and $\omega$ $\Sigma_u^*$ with $x, y, z, \omega \in \mathcal{Z}$, then to match this symmetry we will need $x + 2y +z + 2 \omega = 1 \bmod 3$, $x + 2y = 0 \bmod 3$, and $z + 2 \omega = 0 \bmod 3$. However, if the latter two equations are true, this implies $x + 2y +z + 2 \omega = 0 \bmod 3$, so it's impossible to satisfy all three equations simultaneously. Therefore, there is no representation we can build out of $\Sigma_{u,d}$ representations that will contribute to any of the blocks mixing the $T$ and $B$ quarks with the SM, $U$, and $D$ quarks.

The corrections to the strong CP phase stemming from a correction to the $T$ and $B$ quark mass matrices are identical to those occurring in our non-renormalizable model, which we have already demonstrated are well within experimental constraints, so we will not reiterate our discussion of these corrections here. We therefore only need to confirm that the mass matrices for the $u$, $d$, $U$, and $D$ quarks do not experience large corrections. To estimate these corrections' magnitude, we therefore consider the block mass matrix for the up-like quarks excluding the $T$ quarks, arriving at
\begin{align}
    \widetilde{\mathcal{M}}^u = \begin{pmatrix}
    \frac{y_u v}{\sqrt{2}} \delta_{11} & \frac{y_u v}{\sqrt{2}}(1 + \delta_{12})\\
    y_U \langle \Sigma_u \rangle + \delta_{21} f_u & (1 + \delta_{22})M_u
    \end{pmatrix},
\end{align}
where $\delta_{ij}$ is a small dimensionless matrix correction to the $(i,j)$ block of the mass matrix---the dimensionful terms multiplying each $\delta_{ij}$ are necessitated by the representations of the quarks (for example, an operator in the $(1,1)$ block must include an insertion of the SM Higgs VEV), or simply by the assumption that the higher-order corrections are subleading to those of the tree-level mass matrix. Treating $\delta$ as an expansion parameter, we have 
\begin{align}
    \det (\widetilde{\mathcal{M}}^u) = -\frac{y_U^3 y_u^3 v^3}{2 \sqrt{2}} \det\langle \Sigma_u \rangle + O(\delta) ~,
\end{align}
the complex phase of which is precisely cancelled by the complex phase of the determinant of the $T$ mass matrix. Generically, the determinant of $\widetilde{\mathcal{M}}^u$ in the presence of the correction terms will be
\begin{align}\label{eq:ren-corrected-mass-matrix}
    \det(\widetilde{M}^u) = \frac{y_u^3 v^3}{2 \sqrt{2}} M_u^3 \det ( 1 + \delta_{22}) \det \bigg( \delta_{11} - \frac{y_U}{M_u} (1 + \delta_{12})(1 + \delta_{22})^{-1}(\langle \Sigma_u \rangle + M_u \delta_{21}) \bigg).
\end{align}
We can now simplify our work somewhat by making several observations. First, we note that both $\delta_{11}$ and $\delta_{21}$ must both be operators which have the \emph{same} flavor group representation as those higher-order operators which contribute to the SM quark mass matrix in the non-renormalizable theory, namely, $(3,1,\overline{3})$ under $SU(3)_Q \times SU(3)_d \times SU(3)_u$ and a $U(1)_u$ charge of $-1$. Hence, we can perform a very similar decomposition to that which we performed in the non-renormalizable theory, being able to write
\begin{align}
    \delta_{11,21} = \mathcal{C}_{11,21} \frac{\langle \Sigma_u \rangle}{f_u} ~,
\end{align}
where $\mathcal{C}_{11,21}$ are defined analogously to the scalar product $\mathcal{C}$ in Eq.~(\ref{eq:GeneralCPViolation}), up to additional flavor group singlet operator insertions that will only result in real rescalings of $\mathcal{C}$. Furthermore, we note that $\delta_{12}$ and $\delta_{22}$ are simply either singlets under the SM gauge and flavor groups, or have the representation $(8,1,1)$ under $SU(3)_Q \times SU(3)_d \times SU(3)_u$ and zero $U(1)_d$ and $U(1)_u$ charge. Since the effect of either such operators on the determinant in Eq.~(\ref{eq:ren-corrected-mass-matrix}) can be absorbed into a redefinition of $\mathcal{C}_{11}$ and $\mathcal{C}_{21}$, we see that the most general correction to the determinant in this equation is
\begin{align}
    \det(\widetilde{\mathcal{M}}^u) = - \frac{y_u^3 y_U^3 v^3}{2 \sqrt{2}} \det \bigg(1 - \frac{M_u}{y_U f_u} \mathcal{C}_{11} + \mathcal{C}_{21} \bigg) \det \langle \Sigma_u \rangle ~.
\end{align}
The correction to the complex phase of $\det(\widetilde{\mathcal{M}}^u)$, therefore, is ultimately of the same form as the corrections to the SM quark mass matrices in the non-renormalizable theory, discussed in the main text and Appendix \ref{sec:more-operators}. Therefore, the correction to this phase (and the corresponding phase for the down-like quark mass matrix) in the renormalizable theory should be of a similarly minute magnitude. Therefore, just as in our non-renormalizable setup, the renormalizable theory should be expected to yield a strong CP phase well within current experimental bounds.

Finally, we can comment briefly on the phenomenological characteristics of the renormalizable theory, versus the non-renormalizable one. After integrating out the heavy quarks $U$ and $D$, the theory should be qualitatively quite similar to the renormalizable construction, (up to possible $O(1)$ modifications of the diagonal values of $R_{d,u}$ necessary to recreate the observed quark masses), and therefore any processes involving solely the SM quarks are left virtually unchanged. Similarly, since the $B$ and $T$ quark sectors are identical in the renormalizable and non-renormalizable setups, the influence of these quarks on, for example, $\Delta N_{\rm eff}$ is identical in the renormalizable theory to the results we have found in the non-renormalizable theory. The sole major difference between the renormalizable and non-renormalizable theory lies in the cosmology of the model when the reheating temperature approaches the masses of the $D$ and $U$ quarks, or approximately $M_{d,u} \sim f_{d,u}$. In this case, we may find substantial additional Goldstone boson production from decays of $D$ and $U$ quarks to SM quarks, which would render various Goldstone bosons more likely to be thermally coupled to the SM bath at these temperatures. However, since reheating temperatures near $f_{d}$ or $f_{u}$ generally already result in large contributions to $\Delta N_{\rm eff}$ from Goldstone couplings to the $B$ and $T$ quarks, some of which are hierarchically lighter than these scales, we can assume that most of the parameter space in which the $D$ and $U$ quarks have a significant effect on $\Delta N_{\rm eff}$ are already disallowed by existing constraints on $\Delta N_{\rm eff}$ in the non-renormalizable theory.

\setlength{\bibsep}{3pt}
\bibliographystyle{JHEP}
\bibliography{scp}

\end{document}